\documentclass[preprint,12pt]{aastex}

\shorttitle{Dense Molecular Gas In A Young Cluster Around
MWC 1080} \shortauthors{Wang et al.}

\newcommand{\msun}{M$_{\sun}$}

\begin{document}

\title{Dense Molecular Gas In A Young Cluster Around
MWC 1080 -- Rule Of The Massive Star}
\author{Shiya Wang, Leslie W. Looney}
\affil{Department of Astronomy, University of Illinois,
1002 W. Green St., Urbana, IL 61801}
\email{swang9@astro.uiuc.edu, lwl@uiuc.edu}

\author{Wolfgang Brandner}
\affil{Max-Planck-Institut f$\ddot{u}$r Astronomie, Heidelberg, Germany}

\author{Laird M. Close}
\affil{Steward Observatory, University of Arizona}

\begin{abstract}
We present CS $J = 2 \rightarrow 1$,
$^{13}$CO $J = 1 \rightarrow 0$, and C$^{18}$O $J = 1 \rightarrow 0$, 
observations with the 10-element Berkeley Illinois Maryland Association (BIMA) 
Array toward the young cluster around the Be star MWC 1080. 
These observations reveal a biconical outflow cavity with size $\sim$
0.3 and 0.05 pc for the semimajor and semiminor axis and
$\sim$ 45$\arcdeg$ position angle.
These transitions trace the dense gas, which is likely the swept-up
gas of the outflow cavity, rather than the remaining natal gas or
the outflow gas. The gas is clumpy; thirty-two clumps are identified.
The identified clumps are approximately gravitationally bound and
consistent with a standard isothermal sphere density,
which suggests that they are likely collapsing protostellar cores.
The gas kinematics suggests that there exists velocity gradients implying 
effects from the inclination of the cavity and MWC 1080. 
The kinematics of dense gas has also been affected by either outflows
or stellar winds from MWC 1080,
and lower-mass clumps are possibly under stronger effects from MWC 1080 
than higher-mass clumps.
In addition, low-mass cluster members tend to be formed in the denser and more turbulent cores,
compared to isolated low-mass star-forming cores.
This results from contributions of nearby forming massive stars,
such as outflows or stellar winds.
Therefore, we conclude that in clusters like the MWC 1080 system, 
effects from massive stars dominate
the star-forming environment in both the kinematics and dynamics
of the natal cloud and the formation of low-mass cluster members. 
This study provides insights into the effects of MWC 1080 on its natal cloud,
and suggests a different low-mass star forming environment in clusters compared to isolated
star formation. 
\end{abstract}


\section{Introduction}

Stars are born deep within molecular clouds.
For the last few decades, it has been well established 
that an isolated solar-like (low-mass) star forms and evolves through 
the process of active infall and 
circumstellar accretion of the material provided by the natal cloud,
before reaching the main sequence stage \citep[e.g.][]{Shu1993}.
At the same time, the forming star also effectively disperses
the surrounding molecular gas via outflows and jets
\citep[see review,][]{Arce2007}.
Molecular outflows have been suggested to be the main sources
clearing the circumstellar material around a forming low mass star.
Highly collimated CO flows are ubiquitous 
with low mass protostars of age $<$ 10$^{5}$ yrs, 
having velocities of 10-100 km s$^{-1}$
and typical mass loss rates of 10$^{-6}$ \msun yr$^{-1}$.
However, there has been increasing evidence showing that
the outflow characteristics, especially the collimation, and
even the mechanisms, actually evolve with the low mass protostar
\citep[e.g.][]{Tobin2007}.
Nevertheless, the star forming environment is dramatically
impacted by the evolution of the forming star.

The formation of an intermediate-mass star (3 - 20 \msun) 
is probably similar to low-mass star formation.
For example, the Herbig Ae/Be stars \citep[HAeBes,][]{Herbig1960} are thought to be
the intermediate-mass classical T-Tauri stars (CTTs) \citep[see review,][]{Waters1998},
which are protostars with primordial accretion disks but 
mostly dispersed infalling envelopes.
However, more and more observational evidence suggests that
the formation of massive stars, especially with spectral types
of early B to O, is not simply a scaled-up version of 
low-mass star formation \citep[see review,][]{Zinnecker2007}.
Nevertheless, like a low mass star, gas dispersal is also expected for a HAeBe.
\citet{Fuente1998a,Fuente2002} mapped the molecular gas around HAeBes
with different evolutionary stages and showed that the material 
around HAeBes are usually continuously
swept out by outflows to form cavities as the HAeBes evolve.
This also implies an alternative classification to characterize 
the evolutionary stage of HAeBes,
in addition to the near-infrared scheme \citep[][]{Hillenbrand1992}. 
Indeed, outflows associated with intermediate mass to massive protostars
have started to be identified \citep[e.g.][]{Wu2005}. It is suggested that the outflows
of B stars have higher mass loss rates of 10$^{-5}$ to 10$^{-3}$ \msun yr$^{-1}$
with less collimation \citep[e.g.][]{Wu2005}, compared to outflows from low-mass protostars.

In addition to the formation of isolated stars, 
more and more studies have started to focus on cluster systems,
which are actually the major mode of star formation in the Galaxy
\citep[][]{Carpenter2000,Lada2003}.
For example, \citet{Ridge2003} mapped the dense gas around young
clusters within 1 kpc and also suggested a similar classification of cloud
evolution, as seen in HAeBe systems \citep[][]{Fuente1998a,Fuente2002}. 
However, as a typical cluster normally contains several intermediate- to
high-mass stars surrounded by 10's to 100's low-mass stars
\citep[e.g.][]{Hillenbrand1995,Wang2007},
one would expect that the forming environment of those low-mass 
cluster members must be dynamically affected by the nearby more massive stars.
Is there any difference or modification of gas dispersal processes in clusters,
compared to isolated stars-- What roles do massive stars play in clusters?
On the other hand, how do low-mass cluster members contribute?

In order to investigate the gas dispersal in the cluster environment,
we study one of the nearby embedded young
clusters around HAeBes, the MWC 1080 system.
The MWC 1080 system is a small stellar group embedded within the 
dark cloud LDN 1238 \citep[][]{Lynds1962} at 2.2 kpc \citep[][]{Canto1984,Abraham2000}. 
The most luminous star, MWC 1080 (V628 Cas), 
has been classified as a B0e star \citep[][]{Cohen1979},
with 20.6 M$_{\odot}$,
$10^4$ L$_{\odot}$ \citep[][]{Hillenbrand1992,Hillenbrand1995},
and a flat optically thick circumstellar disk
(Hillenbrand Class I object for Herbig Ae/Be stars).
Within 0.2 pc, 
it is surrounded by at least 15 stars presenting large near infrared excesses \citep[][]{Hillenbrand1995}.
Not only infrared excesses, 
but the discoveries of molecular outflows, 
a P-Cygni feature in the H$\alpha$ line, and several nearby HH objects, 
also indicate active accretion 
and thus the youthfulness of this system ($<$ 1 Myr) 
\citep[][]{Fuente1998a,Finkenzeller1984,Poetzel1992,Yoshida1992}. 
\citet{Levreault1988} estimates an outflow age of 2.2 $\times$ 10$^{5}$ yrs.
Thus, the age of this system is $\sim$ 0.1 - 1 Myr.
Single-dish maps of CO, $^{13}$CO, C$^{18}$O, and CS also show that 
this group is associated with more than 1000 M$_{\odot}$ of molecular material within 1 pc
\citep[][]{Hillenbrand1995} and $\sim$ 10 M$_{\odot}$ within 0.08 pc \citep[][]{Fuente2002}.
Judged by the age of MWC 1080, this system is actually young enough to still have
molecular gas but also old enough to have shown the influence of MWC 1080--
the natal gas is dispersing.
Therefore, it provides a valuable system to study
the molecular gas evolution and star formation in a cluster environment.

In this paper, we present high spatial- and spectral-resolution data obtained with
the BIMA array in the CS(2-1), $^{13}$CO(1-0), C$^{18}$O(1-0),
and 3 mm dust continuum emission toward the MWC 1080 system.
BIMA's high angular resolution can probe the dense gas and dust 
in this young stellar cluster on spatial scales of
1000's to 10,000's of AU.
These molecular transitions are typical tracers for dense gas due to their
high critical densities ($n\,\sim\,10^4\,-\,10^{6}$ cm$^{-3}$),
which helps reveal star-forming clumps and further probe the kinematics and
physical conditions in the system.

\section{Observations}

MWC 1080 was observed using the line transitions CS $J = 2 \rightarrow 1$,
$^{13}$CO $J = 1 \rightarrow 0$, and C$^{18}$O $J = 1 \rightarrow 0$ 
with the 10-element Berkeley Illinois Maryland Association (BIMA) 
Array \citep[][]{Welch1996}.
The CS $J = 2 \rightarrow 1$ ($\nu$ $=$ 97.981 GHz) observations in
C and B array configurations were obtained in October and December 2003.
The correlator was configured with the line window of a velocity range of
76 km s$^{-1}$ with 0.3 km s$^{-1}$ per channel, and two 600 MHz bands
for continuum.
The system temperatures during the observations ranged from 150 - 700 K.
The $^{13}$CO $J = 1 \rightarrow 0$ ($\nu$ $=$ 110.201 GHz) and
C$^{18}$O $J = 1 \rightarrow 0$ ($\nu$ $=$ 109.782 GHz) observations in
B and C array configurations were obtained in March and April 2004.
The correlator was configured with the line window of a velocity range of
135 km s$^{-1}$ with 1.0 km s$^{-1}$ per channel, and two 150 MHz bands
for continuum.
The system temperatures during the observations ranged from 230 - 1000 K.

The data were reduced with the MIRIAD package \citep[][]{Sault1995} and 
mapped using two array configurations with various $u$, $v$ weighting schemes
to stress structures on spatial scales from 2$\arcsec$ to 9$\arcsec$.

We also obtained a continuum submillimeter map of MWC 1080 
at $\lambda\,=$ 850 $\mu$m from the archive data of the Submillimeter
Common User Bolometer Array (SCUBA) instrument on the James Clerk
Maxwell Telescope (JCMT).  

\section{The Morphology of Dense Gas around MWC 1080}

The BIMA interferometry data provide the highest resolution 
observations to date, revealing the dense gas in this cluster. 
We have weighted the data to probe the structures with
spatial scales of 2$\arcsec$ - 9$\arcsec$ for
CS, $^{13}$CO, and C$^{18}$O.
These sizes correspond to spatial resolutions roughly from
20000 AU down to 4000 AU, which is a good scale of probing
low-mass star forming clumps \citep[e.g.][]{Looney2000}. 

\subsection{The Distribution of Dense Gas}

The left portion of Figure 1 plots the velocity-integrated CS contours overlaid 
on an K$\arcmin$-band adaptive optics observation from the 3.6m CFHT 
of the core stellar population
(Wang et al. in prep., Paper II hereafter). The right portion zooms into the core with
this $\sim$ 0.1$\arcsec$ resolution K$\arcmin$-band image 
(Paper II), which illustrates not only the distribution of cluster members but also 
the morphology of a reflection nebulae with a hourglass structure
at the NW-SE direction (called the hourglass axis).
This Hokupa'a 36 \citep[][]{Graves1998} adaptive-optics near-infrared (JHK$\arcmin$) observation and
the photometry of cluster members will be presented in Paper II.
The reflected light results from the existence of dense dust whose surface is
illuminated by the UV radiation from MWC 1080.

The CS map shows two distinct branches of emission (Figure 1),
which we call East and West for the left and right branch,
respectively. 
Figure 1 shows that the distribution of CS emission is clearly aligned with
the obscuring and scattering dust seen in the near-infrared map.
This suggests the existence of a previous bipolar outflow 
that cleaned out the natal material along the direction perpendicular 
to the hourglass axis (the outflow axis), 
and helped clear the view to the stellar population along this direction. 
However, the CS emission is only associated with the upper part of the dense dust,
revealing only the upper half of the cavity structure;
the molecular gas is denser in the upper part than the lower part.

Moreover, 
the $^{13}$CO map also supports the suggestion of bipolar outflows 
by revealing a nearly complete biconical cavity around MWC 1080 (see Figure 2).
Figure 2 plots the velocity-integrated $^{13}$CO and C$^{18}$O contours
as well as the $\lambda$ $=$ 850 $\mu$m and $\lambda$ $=$ 3 mm dust continuum.
The $^{13}$CO morphology
also gives an lower limit of the estimation for the outflow cavity with
$\sim$ 0.3 and 0.05 pc for the semimajor and semiminor axis and
$\sim$ 45$\arcdeg$ position angle.
This is just a lower limit because possible inclination or projection effects
have not been taken into account.
\citet{Fuente1998a,Fuente2002} investigates the evolution of 
dense gas dispersal around HAeBes by presenting IRAM single-dish $^{13}$CO
and CS observations. They propose an evolutionary sequence
from stars that are still embedded within dense clumps and 
associated with bipolar outflows (Type I) to stars that have dispersed their
natal clouds and formed cavities (Type III). 
MWC 1080 is shown to be their Type II system.
Our high spatial-resolution observations further show that
although this system is still young, age $<$ 1 Myr,
the bipolar outflows in the MWC 1080 system has already constructed 
a small cavity, which is revealed for the first time.
The size of this cavity is much smaller, compared to the cavities 
seen in older systems, such as HD 200775, a HAeBe with age of 8 Myr 
and a cavity of size $\sim$ 1.5 pc $\times$ 0.8 pc \citep[][]{Fuente1998b}.
This might suggest that the bipolar outflow activity around MWC 1080
is still an ongoing process that will form a larger cavity.

From Figure 2, the dust morphology at $\lambda$ $=$ 850 $\mu$m is similar to 
the morphology of dense gas, which infers that the dense gas 
traces well the dust around MWC 1080.
The high resolution of the $\lambda$ $=$ 3 mm continuum BIMA map reveals five dust clumps,
which are consistent with dense gas distribution as well.

In addition to the morphology, more information, such as the clumpiness,
kinematics and physical conditions inside this system, 
can be further investigated by extracting the mass and velocity distribution
of dense gas. Studying the clumpiness of a molecular cloud is essential as
it is usually closely related to the fragmentation and 
collapsing of molecular cores inside the cloud, 
which provides valuable insights to the ongoing star-forming activities 
in the system. On the other hand, the kinematics of gas and physical conditions
reveal the natal environment where stars were born.
As CS traces a denser region than $^{13}$CO, we mainly use CS to examine the
clumpiness of dense gas, and use both CS and $^{13}$CO to reveal
the kinematics and physical conditions in this system.

\subsection{Clumpiness of Dense Gas}

Clumpiness is always observed in molecular clouds.
This is not surprising as a star forms through collapsing the natal cloud
after reaching certain collapsing criteria. Therefore, local fragmentation
and further clumping are naturally two typical characteristics 
in star forming clouds.
With the ability of interferometric observations to peer into dense
layers of the molecular cloud, we can study how clumpy the dense gas is and
its clump size, compared to those cores forming isolated stars.

A clump is defined as a local density enhancement, which results in a
flux density enhancement observationally.
A molecular cloud without clumpy structures will not show multiple local emission
peaks in its map, no matter how small the probing scale, or beam size.
On the other hand, if clumps exist in a molecular cloud,
the number of local emission peaks will increase with decreasing scale,
until all clumps have been picked up or some clumps are too weak to be detected.
Therefore, high resolution maps are essential to distinguish clumpy structures in 
the molecular clouds.

Figure 3 plots the CS contours at four different mapping scales with
8$\arcsec$, 4$\arcsec$, 3$\arcsec$, and 2$\arcsec$ beams.
\citet{Fuente1998a} published the IRAM single-dish molecular line 
and continuum data of MWC 1080 and showed one emission peak 
near MWC 1080 in their $^{13}$CO (1-0) 
and CS (3-2) maps with a 24 $\arcsec$ and 16 $\arcsec$ beam respectively, 
and two emission peaks around MWC 1080 at the 1.3 mm continuum map 
with a $\sim$11$\arcsec$ beam.
By comparing these maps, more and more clumps are seen from
their maps to our Figure 3, and from Figure 3(a) to 3(c).
This indicates that the dense gas around MWC 1080 is not a single
big cloud centered on MWC 1080 but rather consists of small clumps,
with clump sizes down to $\sim$ 3$\arcsec$ (6600 AU).
However, there is not much difference in the clumpy structures between 
Figure 3(c) to Figure 3(d), which suggests that the $u$, $v$ weights used in 
Figure 3(d) are less suitable to the majority of the dense gas.
The reason is that the dense gas in Figure 3(d) is resolved out with the
interferometer. 
Therefore, in this paper, we will use Figure 3(c) to further identify clumps
and study their physical properties.

The main purpose of identifying clumps is to obtain individual 
star-forming building blocks that are capable of forming one or multiple
low-mass protostars, in order to investigate their physical condition,
such as mass, velocity dispersion, spectral feature, etc., in each system.
In order to identify clumps, we basically adapt the idea of 
the ${\it Clumpfind}$ method \citep[][]{Williams1994} to:
(1) select local peaks with fluxes above 3 $\sigma$ in the velocity-integrated 
CS map (Figure 3(c)), 
(2) trace down to the half-maximum (HM) flux level of contours, 
which determines the sizes of clumps by fitting an ellipse,
(3) assign the contour to the nearest
peak once this contour is shared by more than one peak,
(4) extract the spectrum and mass of each clump within the clump size, and
(5) fit the extracted spectra with Gaussian to obtain the peak flux,
peak velocity, and line width (FWHM).
This is different from directly using three-dimension cubes because
it can avoid some specific star forming features, such as a P-Cyg profile being
mis-separated into two clumps.

Table 1 lists parameters of 32 identified clumps, including
coordinates, sizes, and integrated fluxes (I$_t$).
The a$_o$, b$_o$, PA$_o$, a, b, and PA are the observed and deconvolved
sizes, including the major axis (a), minor axis (b), and the position angle (PA),
respectively. We directly use the beam size as the upper limit of 
clumps with deconvolved sizes smaller than the beam size.
The R$_{eq}$ is $\sqrt{ab}/2$, the equivalent size 
assuming a spherical clump.
The sizes of identified clumps range from $\sim$ 4000 AU to 10000 AU.
This is generally consistent with a typical system forming a 
single isolated low mass star \citep[e.g., $\sim$ 5000 AU,][]{Looney2003}.

Moreover, the eccentricity of clumps, $\varepsilon$, 
derived by a and b, can be used to characterize the clump shape.
From the derived values of projected eccentricity, we can see that
not all clumps have circular morphology, especially for those closer to
MWC 1080, which show somewhat elongated structures 
with larger eccentricities. 
In order to examine the spatial distribution of clump eccentricity,
we plot the eccentricity vs. the distance to MWC 1080 (Figure 4(a))
and to the outflow axis (Figure 4(b)) for those resolved clumps.
This plot shows that clumps closer to both the outflow axis and 
MWC 1080 tend to have larger projected eccentricities.
This implies that clumps are more elongated when located closer to MWC 1080.
This also suggests possible effects from outflows on
the morphology of dense gas, especially modifying the shape of clumps.

\section{The Kinematics of Dense Gas}

We study the kinematics of the dense gas 
by investigating its velocity distribution.
The spectral resolution of CS, $^{13}$CO, and
C$^{18}$O, observations are 
0.299, 1.034, and 1.034 km s$^{-1}$, respectively,
with V$_{LSR}$ at 29.3 km s$^{-1}$ \citep[][]{Hillenbrand1995}. 
As can be seen in Figure 1 and 2, CS and $^{13}$CO trace
more gas than C$^{18}$O does and CS traces gas with higher density
than the other two transitions. We will use both
CS and $^{13}$CO to display the overall velocity distribution 
of dense gas and specifically use CS to examine
the kinematics of identified clumps.

\subsection{Overall Velocity Distribution}

Figure 5 displays the integrated spectra 
of CS, $^{13}$CO, and C$^{18}$O emission, and
Figure 6 shows the integrated CS spectra of East and West.
From Figure 5, we can see that all CS and C$^{18}$O 
and most $^{13}$CO emission are blue-shifted, 
compared to the V$_{LSR}$. 
In addition, a double-peaked feature is seen in the integrated 
spectra of $^{13}$CO and CS.
Figure 6 shows that
the spectrum of East is single-peaked with
a broad linewidth, while West contains two components.
Table 2 lists the Gaussian-fitted parameters of these
integrated spectra. 
From all integrated spectra, a large velocity gradient of the dense gas 
is seen throughout the whole system;
the separation of peak velocity between two 
components in West is as large as 3.3 km s$^{-1}$.

In order to reveal the velocity distribution and investigate the double-peaked feature,
Figure 7 shows the position-velocity (PV) diagrams for East and West
from CS and $^{13}$CO maps.
This is made by applying two cuts from south to north toward both portions,
as the dense gas of two portions are elongated along the outflow axis.
Interestingly, it shows that two components seen in the integrated CS
spectrum of West are also spatially distinguishable--
the blue and red component in the spectrum are related to
the northern and southern part of the dense gas in West, respectively.
\citet{Canto1984} also detected a double-peak feature 
on the spectrum of dense gas around MWC 1080, but explained it 
as the results from self-absorption due to 
poor spatial resolution. 
Our high-resolution interferometric data shows that 
this double-peaked feature
actually comes from two sets of gas with different velocities.
In addition to West, Figure 7 also shows that 
there is a systematic velocity gradient in East
from -35 km s$^{-1}$ (bluer, south) to -30 km s$^{-1}$ (redder, north),
especially seen in the $^{13}$CO map.
This gradient may result from an inclination of the outflow cavity,
as the direction of the gradient is along the outflow axis.

Not only is there a velocity variation along the outflow axis,
there is also a velocity gradient along the hourglass axis,
which is perpendicular to the outflow axis.
Figure 8 is the PV diagrams of the CS map with three cuts 
perpendicular to the outflow axis and moving away from MWC 1080.
This also shows that the velocity dispersion along the farther cut is larger
than that along the closer cut from MWC 1080.
This is contrary to the idea that outflows or stellar winds from MWC 1080
contribute to the nonthermal motion, thus increase the velocity dispersion, 
of gas closer to MWC 1080.

\subsection{Spectra of Identified CS Clumps}

Figure 9 displays example spectra of our identified clumps.
Table 3 lists Gaussian-fitted parameters of the clumps
that can be fitted with Gaussian.
A few of them show P-Cyg-like absorption; several of them contain
two peaks, which might come from either self-absorption or
a secondary component (another clump along the line of sight).

\section{Estimating the Column Density and Mass of Dense Gas}

Several methods have been used to estimate the column density from 
molecular emission, such as a direct estimation from individual transitions
\citep[][]{Miao1995,Friedel2005}, the Large Velocity Gradient approximation
\citep[LVG,][]{Goldreich1974}, rotational temperature diagrams \citep[RTD, e.g.,][]{Friedel2005}, etc.
Since we only have individual transitions for each molecular species,
we use a simple LTE method for all column density estimated in this paper,
and also use the LVG models for those CS clumps with fitted spectra
for comparison.

\subsection{LTE Approximation}

By assuming LTE with an excitation temperature T$_{ex}$,
small opacity and negligible background continuum contribution, 
the total column density from a single transition using an array 
can be derived by \citep[e.g.,][]{Miao1995},

\begin{equation}
N = \frac{2.04IC_{\tau}}{\theta_a\theta_bS\mu^2\nu^3}Qe^{\frac{E_u}{T_{ex}}}\,\times\,10^{20}\,cm^{-2}.
\end{equation}
I is the total integrated intensity in Jy beam$^{-1}$ km s$^{-1}$, 
$\theta_a$ and $\theta_b$ are the FWHM sizes of beams in arcseconds, 
S, $\mu^2$, $\nu$, Q, and E$_u$ are the line strength, dipole moment in Debyes, 
line frequency in GHz, partition function, 
and energy in K of the upper state, respectively (Table 4).
We use the deconvolved source sizes, a and b, as $\theta_a$ and $\theta_b$ here.
C$_{\tau}$ is the opacity correction factor \citep[][]{Goldsmith1999},
\[
C_{\tau} = \frac{\tau}{1-e^{-\tau}}.
\]
The optical depth, $\tau$, can be derived by \citep[][]{Rohlfs2000},
\[
\tau = -ln[1-\frac{T_{MB}}{J(T_{ex})-J(2.73)}],
\]
where T$_{MB}$ is the main beam brightness from observations and
J(T) is given by $\frac{h\mu}{k}\,\frac{1}{e^{h\mu/kT}-1}$.
Table 1 lists some of these parameters for CS(2-1), $^{13}$CO(1-0), and
C$^{18}$O(1-0) \citep[][]{Rohlfs2000}.

With estimated column density, mass can be obtained by assuming the abundance ratio
to H$_2$, X$_{CS}$, X$_{^{13}CO}$, and X$_{C^{18}O}$ $=$ 10$^{-9}$, 1.26 $\times$ 10$^{6}$, 
and 1.7 $\times$ 10$^{-7}$, respectively \citep[][]{Rohlfs2000}. 
An excitation temperature, T$_{ex}$ $=$ 20 K, is assumed, 
as a medium value often seen in dense cores forming massive stars \citep[see review,][]{Zinnecker2007}.
Therefore, the mass can be derived by

\begin{equation}
M = \mu m_H \frac{N}{X}(1.133abD^2),
\end{equation}
where $\mu$ $=$ 2.33 is the mean molecular weight, m$_H$ is the mass of a hydrogen atom,
and D is the distance \citep[$=$ 2.2 kpc for MWC 1080;][]{Canto1984,Abraham2000}. 

The estimated CS column densities and masses for CS clumps are listed in Table 5, 
labeled as M$_{LTE}$, along with the estimated optical depth. 
The masses of these clumps range from $\sim$ 1 - 10 M$_{\odot}$,
with mean optical depth $\tau$ less than 1.
We also estimate the maximum optical depth based on the peak
emission of the CS map, which gives optical depth $\sim$ 1.2 in this system.

Using the same assumptions and eq. (1) and (2), the total masses of dense 
gas are also calculated.
There are $\sim$ 800, 1000, and 900 M$_{\odot}$ for CS, $^{13}$CO, 
and C$^{18}$O, respectively,
within the area of 0.7 pc radius from the MWC 1080.

\subsection{LVG Model}

Another method, that is frequently used especially for optical thick lines, is the Large
Velocity Gradient (LVG) approximation \citep[][]{Goldreich1974}.
This is a radiative transfer model that takes 
optical depths into accounts for photon transport. It basically assumes a cloud with
a systematic velocity gradient with velocity increasing away from the cloud center, so that
a local treatment for photon transport can be approximated. We also estimate
column densities and masses of CS clumps based on the LVG model, in order to
compare to the previous LTE calculation. 
We only apply LVG for those 14 CS clumps with well fitted spectral information from Table 3.

We use the $\it{lvg}$ task in MIRIAD to generate grids of the LVG model with
kinetic temperatures of 10 - 30 K, number densities of 10$^{4}$-10$^{8}$ cm$^{-3}$, 
and column densities of 10$^{13}$-10$^{15}$ cm$^{-2}$. 
By using these grids to fit the observed brightness temperature,
a range of CS column density can be found that can produce the observed emission.
We further narrow the obtained range of CS column density by assuming 
the optical depth $\tau$ $\lesssim$ 2. As there is a lack of observations of
NH$_3$ to further constraint the kinetic temperature, we simply use the range of
10 - 30 K, which is typically used for both low-mass and massive star forming regions. 
The derived column densities and masses of CS clumps are listed in the Table 5.

We also compare the estimation of M$_{LTE}$ and M$_{LVG}$ for these clumps.
All clumps, except clumps A2 and A4, show a roughly good agreement 
between the derived M$_{LTE}$ and M$_{LVG}$. 
This consistency suggests that most CS clumps
are under good LTE approximation with small opacity.
Clumps A2 and A4 have the brightest peak emission. 
The reason why the LVG model gives much larger column density than the LTE approximation
is that there is either a larger opacity or a higher temperature inside these two clumps.
Large opacity will give a larger column density in the LTE approximation; while higher temperature
than 30 K will allow smaller column density in the LVG model to produce lines 
with such strong intensity.
Both are reasonable explanations, especially the temperature in these two clumps is likely
higher than usual, as they are located very close to the source of strong UV radiation, MWC 1080.
In this paper, we will use the M$_{LTE}$ for further discussion.

\subsection{Virial Mass}

With the estimated masses and the velocity dispersion of clumps, we can further study the virial
condition inside each clump by comparing clump mass with the virial mass, M$_{vir}$.
M$_{vir}$ is the mass when the system is in virial equilibrium, when 
the time
average over the kinetic energy is equal to half of the potential energy.
In this situation, the system is gravitationally bound.

We assume a spherical symmetric clump with total mass M, radius R, and
a density profile, $\rho(r)\,=\rho_0\,r^{-\alpha}$, 
where $\rho_0$ is the central density and r is the radial distance from the clump center.
Therefore,
\begin{equation}
M_{vir} = \frac{5-2\alpha}{3-\alpha} \frac{3R}{G} \frac{\Delta V^2}{8ln2},
\end{equation}
where $\alpha$ $\neq$ 3. Either uniform density or isothermal condition is often assumed while deriving
the virial mass of a cloud. In these two cases, 
\begin{equation}
M_{vir} = \frac{5}{3} \frac{3R}{G} \frac{\Delta V^2}{8ln2}
\end{equation}
for uniform density, and,
\begin{equation}
M_{vir} = \frac{3R}{G} \frac{\Delta V^2}{8ln2}
\end{equation}
for the isothermal cloud with density $\rho(r)\,=\rho_0\,r^{-2}$.

In order to understand the density profile in our system,
we investigate the relation between the total mass and size of CS clumps,
by plotting M$_{LTE}$ and R$_{eq}$ (Figure 10), as used in \citet{Saito2006}.
We obtain a linear relation between these parameters via least-square fittings,
\begin{equation}
log M_{LTE}(M_{\odot}) = (0.87 \pm 0.32) log R_{eq}(pc) + (1.82 \pm 0.67),
\end{equation}
with $\chi_r^2$ $=$ 1.35.
This assumes that all clumps in this system around MWC 1080 have the same density
structure. 
Since $\rho(r)\,=\rho_0\,r^{-\alpha}$, which gives 
$M\,=\,\frac{4\pi\rho_0}{3-\alpha}\,R^{3-\alpha}$, then 
$\rho_0$ $=$ 4.57 $\pm$ 1.69 M$_{\odot}$ pc$^{-3}$ and
$\alpha$ $=$ 2.13 $\pm$ 0.32 are obtained.
Therefore, in our case
\begin{equation}
M_{vir} \sim \frac{3R}{G} \frac{\Delta V^2}{8ln2},
\end{equation}
shows that our systems are consistent with a standard isothermal sphere
density \citep[][]{Shu1977}.

Given R$_{eq}$ and FWHM from Table 1 and 3 as R and $\Delta V^2$,
M$_{vir}$ is calculated and listed in Table 5. 
M$_{vir}$ ranges from $\sim$ 1 - 10 M$_{\odot}$.
By comparing M$_{vir}$ with M$_{LTE}$,
we find that the M$_{vir}$ is $\sim$ 0.4 - 2 times the M$_{LTE}$,
except A8, C1, and C5, the farthest three clumps away from MWC 1080, 
which are 2 - 4 times larger than M$_{LTE}$.
In general, we can conclude that the virial mass is similar to the clump mass,
as the mass estimation usually has an uncertainty to an order of 2.
This is similar to low-mass star forming cores, rather than massive cores
\citep[e.g.,][]{Saito2006}.

\section{Discussion}

This paper presents high resolution data for dense gas in the cluster
environment around the HAeBe star MWC 1080.  The CS transition is
especially used to peer into the dense core region.  Freeze-out depletion
of CS or CO onto dust grains would affect our results.  However, molecular
depletion mostly occurs in colder cores. For example, CS and CO are often
seen to freeze-out onto dust grains in starless cores, which have lower
average temperatures, $\sim$ 10 K \citep[see review,][]{DiFrancesco2007}.

With high resolution, the importance of this study is
being able to reveal the physical conditions 
and examine the sub-structures (i.e. clumps, that may be forming
single low-mass cluster members) within this
cluster-forming cloud, and distinguish
the effects of MWC 1080 and its outflows from
the initial cloud environment.
We will focus on discussing the effects from MWC 1080 on its natal cloud by
characterizing the dense gas, its dispersal history, and
star-forming clumps in this cluster.
To best study star formation in cluster environments would require
a large sample of objects like MWC 1080;
however, these data still provide valuable insights into such systems.

\subsection{Gas Removal in Clusters}

Outflows are the primary sources to remove the natal
material during the formation of stars, in addition to
stellar winds and stellar radiation in the core of starburst clusters.
However, the gas removal process via outflows in clusters has not been 
clearly characterized. 
As there are both high-mass and low-mass members forming
in a cluster, the gas removal process might be different from that in 
a single star forming system.
Especially, does the most massive star still dominate the outflows
that results in similar gas removal processes as seen
from an isolated star?
What is the role of low-mass cluster members in the
strength and collimation of outflows?

In fact, the existence of a bipolar outflow in this cluster around MWC 1080
has been suggested based on its blue and red components
on CO spectra in previous single-dish studies \citep[e.g.,][]{Canto1984,Yoshida1991}.
However, due to poor angular resolution
of single-dish observations, the outflow properties, such as the direction
of the outflow axis, have not been well characterized. 
From our data, a biconical cavity is distinctly revealed
around the most massive star in this system, MWC 1080,
suggesting that the bipolar outflow has been
dominated by MWC 1080.
This also implies evidence of the gas removal by
outflows in clusters. With the observed outflow cavity, the outflow axis
is also clearly defined to have a PA of 45 degree.

Our data also show that this outflow cavity has a size of 0.3 $\times$ 0.05 pc. 
This roughly gives an opening angle of 20 degree
and a collimation factor of 3.
This also means that the outflow activity has begun and formed a small cavity
at the age $<$ 1 Myrs for a HAeBe star like MWC 1080.
However, gas removal via outflows is still an ongoing process in this 
system, as a larger outflow cavity has been seen in a similar 
but older system \citep[HD 200775, 8 Myrs,][]{Fuente1998b}.
A future $^{12}$CO high resolution observation will be helpful to confirm the ongoing outflow
activity in this system, as $^{12}$CO can trace low density outflow gas. 
Assuming a homogeneous mass distribution of gas in the initial natal cloud
and an initial mass density the same as the remaining gas,
we can estimate that there was $\sim$ 1000 M$_{\odot}$ of gas 
in the cavity before outflows removed it.
Therefore, an outflow mass loss rate $>$ 10$^{-3}$ M$_{\odot}/$yr is required
to form this cavity, either assuming a outflow age of $\sim$ 0.2 Myrs 
\citep[][]{Levreault1988} or an age $<$ 1 Myrs \citep[][]{Fuente1998a}.
This value is higher than the typical mass loss rate of outflows from
a single OB star (10$^{-3}$ $-$ 10$^{-4}$ M$_{\odot}/$yr).
In fact, previous CO observations, which trace outflow gas at 
larger scale than this study, have estimated an outflow mass loss rate
of $\sim$ 10$^{-4}$ M$_{\odot}/$yr  
\citep[e.g.][]{Yoshida1991,Canto1984,Levreault1988}.
One explanation of possible higher mass loss rate for CS gas is 
that the outflows from other cluster members are
strong enough to strengthen the overall effects, but not enough
to change the outflow direction, at the scale of our study.
Another possibility is that the remaining gas does not just 
include the natal material but also the swept-up gas by outflows,
so the mass loss rate is overestimated.

In order to clarify whether the observed dense gas
is the remaining gas from the initial cluster-forming cloud
or the swept-up gas by outflows, we further inspect the possible 
non-thermal contribution from outflows in dense gas next.

\subsection{Non-thermal Contributions in Dense Gas}

Molecular clouds, except for single star-forming prestellar cores, 
are generally observed to have broader linewidths
than those caused by the thermal motion from their thermal temperatures.
This is suggested to be resulting from the presence of turbulence
\citep[see review by][]{Mac2004}.
Is this also true for single star-forming clumps in a cluster around
an intermediate- to high-mass star, like the MWC 1080 system?
What is the role of the cluster environment in the kinematics of dense gas
and clumps within one cluster-forming cloud?

Our data show that the linewidth of clumps range from 0.85 to
2.48 km s$^{-1}$, which is much larger than 0.63 km s$^{-1}$,
the linewidth of thermal broadening at 20 K.
If only thermal motion contributes to the linewidth,
the kinematic temperatures of these clumps would range 
from 36 - 310 K. This is unlikely,
except for those clumps very close to MWC 1080,
which heats its surroundings via UV radiation.
Therefore, our results suggest that most gas in this system has a 
non-thermal contribution to their kinetic motions.
However, what contributes to these non-thermal motions?
Initial turbulence, outflows from forming stars, and 
stellar winds from massive stars are possible candidates.

Initial turbulence is the general non-thermal random motion caused by
turbulence in the molecular cloud before forming stars.
\citet{Larson1979,Larson1981} found the linewidth-size relation
for molecular clouds with different scales.
This relation indicates that there is a power-law relation
between the velocity dispersion of molecular clouds and
cloud size. \citet{Caselli1995} further suggested
that the non-thermal linewidth-size relation follows different
power-law trends for low-mass and massive cores.
In other words, non-thermal motion still dominates at small scales
for massive clumps, while low-mass clumps show only thermal motions
at these scales. 
Nevertheless, there exist non-thermal components in the kinematics
of molecular clouds, which provides the initial turbulence.
Based on our data, we roughly estimate that the radius of
this cluster-forming cloud is $\sim$ 0.7 pc.
This gives an estimation for the initial turbulence with linewidth 
of $\sim$ 1 km s$^{-1}$ in our system, assuming the linewidth-size relation 
from \citet{Larson1981}.
This value is smaller than the linewidth from our integrated spectra (Table 3),
which implies that the initial turbulence may not be the only contributor
to the non-thermal motion in this system.

In order to confirm more carefully whether there are extra contributors 
to broaden the linewidth of clumps, we plot the linewidth-size relation
for identified clumps in Figure 11, compared to \citet{Caselli1995}.
\citet{Caselli1995} shows the nonthermal linewidth-size relations for massive
and low-mass cores. Therefore, in this figure, we add the thermal components to
their relations with thermal temperature of 18 K and 10 K for massive (solid line) 
and low-mass (dashed line) cores, respectively, as assumed in \citet{Caselli1995}.
The dotted line indicates the thermal motion of 20 K, which is well
below the linewidths of all clumps.
It also shows that there is either a flat relation or no correlation between
the linewidth and the clump size. When using the relation from 
\citet{Caselli1995} as expected initial turbulence contribution,
this figure suggests that extra contributions dominate over the
initial turbulence and affect more the smaller clumps.
Not only the relation, but the linewidth values of identified clumps
are also overall larger than those in \citet{Caselli1995}.
In addition, Figure 11 also plots the relation between the linewidth 
and the LTE mass of clumps, compared to the result shown in \citet{Larson1981}.
It shows that the low-mass clumps seem to deviate more from Larson's relation.
This indicates that the lower-mass clumps are possibly affected more, which
implies that the extra contributor to the non-thermal motions may come from
a common source outside these clumps, instead of
heated contribution from 
embedded forming stars inside individual clumps.

Both outflows or stellar winds can contribute to the observed non-thermal
components of linewidth. From the Figure 11, those extra contributors 
need to broaden the linewidth by $\sim$ 1-1.5 km s$^{-1}$.
Furthermore, Figure 12 plots the relations for the linewidth 
and mass vs. the distance to MWC 1080.
From the linewidth-distance relation, no correlation is 
shown. 
As outflows or stellar winds from MWC 1080 are possible contributors
to the non-thermal motions of natal gas, the affected linewidth are 
expected to be broader for clumps closer to MWC 1080.
Therefore, this plot possibly suggests that
there is less initial turbulence deep inside the core region of this
cluster-forming cloud.
Besides, it is shown in the linewidth-mass relation that
more massive clumps form closer to
MWC 1080, which is or is close to the center of this cluster-forming cloud.

In short, the kinematics of dense gas in this cluster has been 
effectively modified by outputs from forming stars, 
such as outflows and stellar winds,
instead of still being sustained by the initial turbulence.
This further suggests that the observed hourglass dense gas is much more likely
to be swept-up gas.

\subsection{Implications from the Dense Gas Morphology}

Who forms first in a cluster-- massive or low-mass stars?
This has been an fundamental but difficult question in the study of cluster formation,
because highly embedded nature of protoclusters, where massive star forms, 
have prevented us from catching the earliest stage of the formation of low-mass cluster members.
However, clarifying this question is essential as it helps understand the star forming
environment in clusters and after all most low-mass stars actually form 
in such cluster environments.
This also plays an important role in studying massive star formation as massive
stars often form in clusters along with many low-mass stars.

Our results have suggested that the observed molecular lines most likely 
trace the swept-up gas on the outflow cavity wall, instead of outflows or
remaining gas with initial density. The kinematics of the swept-up gas
has also been effectively modified by outflows or stellar winds from MWC 1080.
The biconical cavity implies a 
domination of MWC 1080 and also eliminates the contribution of 
other cluster members. In addition, Paper II also presents that there 
are $\sim$ 50 low-mass stars within 0.3 pc radius around MWC 1080.
Most of these stars are located inside the outflow cavity (also see Figure 1 and 2).

These results may imply that it is unlikely to form low-mass cluster members after
MWC 1080 in this system. If low-mass cluster members form after MWC 1080 does, 
then the gas dispersal from MWC 1080 should eliminate the formation 
of low-mass cluster members along the outflow direction. 
Hence, the stellar density in the cavity should have been lower than observed,
if these low-mass stars formed after MWC 1080.

However, one puzzle remains about the morphology-- why does CS emission only trace the upper half
of the outflow cavity? In other words, why is the gas in the upper part denser than the lower part?
One explanation is an inhomogeneous distribution of initial natal gas;
the gas is just denser in the upper side than the lower side. 
Another explanation is that this is due to the inclination of outflow cavity, 
as all CS emission is blue-shifted and a gradient has been seen in Figure 8.
The asymmetry of the opening angle between both sides from the $^{13}$CO map also 
may also result from the inclination.

\subsection{Clumpiness of Dense Gas vs. Gas Dispersal History}

As stars form, they disperse their natal material,
so studying the morphology of molecular gas around newly forming stars
has been very helpful to distinguish stars with different evolutionary 
stages \citep[e.g.,][]{Fuente1998a,Fuente2002}.
A star-forming cloud typically evolves from
compact dense gas centered on the star, to dense gas with bipolar outflows,
then finally to a cavity with little gas left.
However, not only the natal molecular cloud, but the clumpiness of dense gas
in the cloud may also be actively affected by the dispersal processes 
from the forming star.
Therefore, the clumpiness of dense gas might be able to provide an alternative
point of view to describe the evolutionary stages of star formation.

The clumpiness of dense gas can be defined as: 
(1) the fraction of mass inside clumps to total mass; or 
(2) the number density of clumps-- number of clumps per projected area.
The first definition is determined by the intrinsic star forming
efficiency in the cloud and the degree of gas dispersal;
the second definition is determined by the intrinsic fragmentation
in the cloud and the degree of gas dispersal.
Both definitions depend on the gas dispersal history because the mass 
inside clumps decreases and the clumps disappear as stars form.
Therefore, 
the clumpiness can actually be used to trace the evolutionary stages of the forming
stars, assuming the same initial conditions in the cloud and similar
star forming processes.
This assumption is valid if we simply apply and compare the clumpiness to
similar systems, for example, to clusters around HAeBes like MWC 1080. 
However, we have to keep in mind that the cloud's initial conditions may dominate
the clumpiness, as different star or cluster forming mechanisms could
result in various initial cloud conditions.

In this paper, we will only use the second definition to discuss the clumpiness. 
This is because the true mass inside clumps can not be fully obtained and
the first definition will be biased with the dependence of mapping scales, 
due to the limitation of interferometry.
\citet{Looney2006} identifies 16 clumps from another cluster around HAeBes
including BD +40$\arcdeg$ 4124. Using the second definition, we find that
the clumpiness in the BD +40$\arcdeg$ 4124 system is $\sim$ 1.6 times larger than
that in the MWC 1080 system.
This suggests that the dense gas seen in the BD +40$\arcdeg$ 4124 system is at a younger 
stage than that in the MWC 1080 system. 

From the infrared SED classification \citep[][]{Hillenbrand1992}, 
BD +40$\arcdeg$ 4124 and MWC 1080
are both classified as Group I objects, which have rising SED slopes.
From the gas dispersal history \citep[][]{Fuente1998a,Fuente2002}, 
they are both associated with compact dense gas, which suggests a similar evolutionary
stage. In addition, from the CS maps using high resolution interferometric 
observations (Figure 1 in Looney et al. 2006 for BD 40$\arcdeg$ +4124 and Figure 5(b) here),
both systems show that the CS distributions are actually offset from the most 
massive stars-- still at a similar evolutionary stage.
\citet{Looney2006} show that the dense gas is around other younger
stars, not BD+40$\arcdeg$ 4124.
The clumpiness discussed here suggests the youthfulness of dense gas 
in the BD +40$\arcdeg$ 4124 system and also shows that the dense gas 
in the MWC 1080 is dominated
by MWC 1080, unlike the BD +40$\arcdeg$ 4124 system.
Indeed, the clumpiness provides an alternative point of view to examine
the evolutionary stages of young stars, other than the morphology of dense gas
or the SEDs of young stars.

\subsection{Dynamics of Clumps}

Our data have shown that
the mass of identified clumps ranges from $\sim$ 0.5 to 10 M$_{\odot}$, 
capable of forming low-mass stars, 
with equivalent radius from 0.01 to 0.04 pc. 
These clumps are, in general, more massive than protostellar cores with similar
sizes. This suggests that clumps formed in a cluster environment 
tend to have higher densities, which is also shown in \citet{Saito2006}.

In addition, all clumps are estimated to have masses similar to the virial masses,
which suggests that they are close to being in virial equilibrium.
Figure 13 compares the LTE mass and LVG mass vs. virial mass for all clumps.
This figure shows that most clumps have masses within 2 times of their virial masses,
which means that they are self-gravitational bound systems.
This is usually seen in low-mass cores; but it
is different from turbulent cores in massive star-forming regions 
\citep[][]{Saito2006}, showing larger virial masses than LTE masses.
This may be due to the external gas pressure helping
support the turbulence \citep[][]{Saito2006}.
However, according to their Table 4, cores with LTE masses $<$ 10 M$_{\odot}$
are actually close to their virial masses,
which is consistent with our results. 

Nevertheless, our clumps in the MWC 1080 system are self-gravitationally bound,
similar to low-mass star-forming cores, but have a higher density like
massive cores. 
This is not surprising as more mass is needed to bound clumps
in order to overcome the non-thermal motions, which are contributed 
not only by initial turbulence but also inputs from massive stars, 
such as outflows and stellar winds.
This does not necessary mean that they will form massive stars 
eventually, because they will experience stronger external gas removal,
such as outflows from nearby massive stars, than those isolated
star-forming cores. 
In fact, we can simply assume that the identified clumps are possible
precursors of those NIR-identified low-mass cluster members (Paper II), 
that are revealed just
because their surrounding gas is strongly dispersed by outflows from MWC 1080.
This means that these clumps will indeed be likely forming low-mass stars.

In addition, from Sec. 5.3, we derive a density profile for the identified
clumps, $\rho(r) \sim r^{-2.13 \pm 0.32}$.
This is consistent with \citet{Saito2006}, which shows a density profile
of $\rho$ $\sim$ r$^{-1.9}$ for cores in massive star-forming regions. 
Our results show that clumps in our system are consistent with 
the density profile of many protostellar collapse models
\citep[e.g.,][]{Tassis2005}.
Therefore, along with the fact that these clumps are gravitationally bound,
the $\sim$ r$^{-2}$ profile from our data suggest that
these clumps are likely collapsing protostellar cores.

We simply conclude that low-mass stars in the cluster environment like
the MWC 1080 system tend to be formed in dense and turbulent cores, 
which are different from isolated low-mass star-forming cores but
similar to massive cores.
However, gas dispersal contribution from the massive cluster member
prevents these dense and turbulent cores from forming massive stars,
instead forming low-mass stars,
which are also revealed earlier than isolated 
low-mass stars are.
This indicates that massive stars in clusters do have effects on the
formation of their low-mass cluster members-- both help and hinder.

\section{Conclusions}

We present BIMA CS(2-1), $^{13}$CO(1-0), C$^{18}$O(1-0), and 3mm continuum observations
toward the young cluster around MWC 1080, which is a $\sim$ 20 M$_{\odot}$ massive
star with the age $<$ 1 Myr.
We summarize our results as follows.

\begin{itemize}
\item A biconical cavity, with size of 0.3 $\times$ 0.05 pc
and $\sim$ 45$\arcdeg$ position angle, is revealed, which suggests the presence of 
bipolar outflows. The outflows are dominated by the MWC 1080, and
effectively modifying the morphology of clumps.
\item The observed molecular lines trace the swept-up gas on the cavity
wall, instead of the initial natal material or the outflow gas.
\item The observed gas is clumpy; 32 CS clumps are identified with mass ranging
from 1 - 10 M$_{\odot}$. All clumps are approximated under the virial equilibrium,
which suggests that they are gravitationally bound,
and isothermal. This suggests that they are likely collapsing protostellar cores.

\item The gas is mostly blue-shifted. We identify two distinct clouds with different
velocities that were thought to be self-absorption. Velocity gradients have also been
revealed suggesting an inclination of the outflow cavity and some effects from MWC 1080.
\item Both overall gas and clumps show broader linewidths than thermal motion at 20 K.
The non-thermal component is possibly contributed by outputs from MWC 1080,
such as outflow and stellar wind, in addition to the initial turbulence
often seen in the molecular cloud.
This suggests that the kinematics of dense gas has been affected by either outflow
or stellar wind from MWC 1080;
lower-mass clumps are more strongly effected from MWC 1080 
than higher-mass clumps.
\item Clumps in clusters have, in general, higher densities than 
isolated star-forming cores. This results from non-thermal contributions,
such as outflows or stellar winds, from nearby forming massive star or stars.
However, low-mass stars can still be forming from these clumps, because
of the increased gas dispersal from MWC 1080.
Therefore, low-mass cluster members tend to be formed in dense and turbulent cores,
which are different from isolated low-mass star-forming cores.
\end{itemize}

In summary, our results show that in the cluster like MWC 1080 system, 
effects from the massive star dominate
the star-forming environment in the cluster, in both kinematics and dynamics
of the natal cloud and the formation of low-mass cluster members.
However, more studies in similar systems like the MWC 1080 cluster are needed in the future,
in order to systematically confirm the effects from massive stars.

\acknowledgements

We thank Dr. Mordecai-Mark Mac Low for valuable suggestions.
We thank Dr. Murad Hamidouche for assistant with the SCUBA data reduction.
S.W. and L.W.L. acknowledge support from the Laboratory for Astronomical Imaging at the
University of Illinois, NSF AST 0228953.
The James Clerk Maxwell Telescope is operated by the Joint Astronomy Centre 
on behalf of the Science and Technology Facilities Council of the United Kingdom, 
the Netherlands Organisation for Scientific Research, and the 
National Research Council of Canada. 
The JCMT Archive project is a collaboration between the Canadian 
Astronomy Data Centre (CADC), Victoria and the James Clerk Maxwell 
Telescope (JCMT), Hilo. 


\begin{thebibliography}{45}
\expandafter\ifx\csname natexlab\endcsname\relax\def\natexlab#1{#1}\fi

\bibitem[{{{\'A}brah{\'a}m} {et~al.}(2000){{\'A}brah{\'a}m}, {Leinert},
  {Burkert}, {Henning}, \& {Lemke}}]{Abraham2000}
{{\'A}brah{\'a}m}, P., {Leinert}, C., {Burkert}, A., {Henning}, T., \& {Lemke},
  D. 2000, \aap, 354, 965

\bibitem[{{Arce} {et~al.}(2007){Arce}, {Shepherd}, {Gueth}, {Lee}, {Bachiller},
  {Rosen}, \& {Beuther}}]{Arce2007}
{Arce}, H.~G., {Shepherd}, D., {Gueth}, F., {Lee}, C.-F., {Bachiller}, R.,
  {Rosen}, A., \& {Beuther}, H. 2007, Protostars and Planets V, 245

\bibitem[{{Canto} {et~al.}(1984){Canto}, {Rodriguez}, {Calvet}, \&
  {Levreault}}]{Canto1984}
{Canto}, J., {Rodriguez}, L.~F., {Calvet}, N., \& {Levreault}, R.~M. 1984,
  \apj, 282, 631

\bibitem[{{Carpenter}(2000)}]{Carpenter2000}
{Carpenter}, J.~M. 2000, \aj, 120, 3139

\bibitem[{{Caselli} \& {Myers}(1995)}]{Caselli1995}
{Caselli}, P. \& {Myers}, P.~C. 1995, \apj, 446, 665

\bibitem[{{Cohen} \& {Kuhi}(1979)}]{Cohen1979}
{Cohen}, M. \& {Kuhi}, L.~V. 1979, \apjs, 41, 743

\bibitem[{{di Francesco} {et~al.}(2007){di Francesco}, {Evans}, {Caselli},
  {Myers}, {Shirley}, {Aikawa}, \& {Tafalla}}]{DiFrancesco2007}
{di Francesco}, J., {Evans}, II, N.~J., {Caselli}, P., {Myers}, P.~C.,
  {Shirley}, Y., {Aikawa}, Y., \& {Tafalla}, M. 2007, in Protostars and Planets
  V, ed. B.~{Reipurth}, D.~{Jewitt}, \& K.~{Keil}, 17--32

\bibitem[{{Finkenzeller} \& {Mundt}(1984)}]{Finkenzeller1984}
{Finkenzeller}, U. \& {Mundt}, R. 1984, \aaps, 55, 109

\bibitem[{{Friedel}(2005)}]{Friedel2005}
{Friedel}, D.~N. 2005, PhD thesis, University of Illinois at Urbana-Champaign,
  United States -- Illinois

\bibitem[{{Fuente} {et~al.}(1998{\natexlab{a}}){Fuente}, {Martin-Pintado},
  {Bachiller}, {Neri}, \& {Palla}}]{Fuente1998a}
{Fuente}, A., {Martin-Pintado}, J., {Bachiller}, R., {Neri}, R., \& {Palla}, F.
  1998{\natexlab{a}}, \aap, 334, 253

\bibitem[{{Fuente} {et~al.}(2002){Fuente}, {Mart{\i}n-Pintado}, {Bachiller},
  {Rodr{\i}guez-Franco}, \& {Palla}}]{Fuente2002}
{Fuente}, A., {Mart{\i}n-Pintado}, J., {Bachiller}, R., {Rodr{\i}guez-Franco},
  A., \& {Palla}, F. 2002, \aap, 387, 977

\bibitem[{{Fuente} {et~al.}(1998{\natexlab{b}}){Fuente}, {Martin-Pintado},
  {Rodriguez-Franco}, \& {Moriarty-Schieven}}]{Fuente1998b}
{Fuente}, A., {Martin-Pintado}, J., {Rodriguez-Franco}, A., \&
  {Moriarty-Schieven}, G.~D. 1998{\natexlab{b}}, \aap, 339, 575

\bibitem[{{Goldreich} \& {Kwan}(1974)}]{Goldreich1974}
{Goldreich}, P. \& {Kwan}, J. 1974, \apj, 189, 441

\bibitem[{{Goldsmith} \& {Langer}(1999)}]{Goldsmith1999}
{Goldsmith}, P.~F. \& {Langer}, W.~D. 1999, \apj, 517, 209

\bibitem[{{Graves} {et~al.}(1998){Graves}, {Northcott}, {Roddier}, {Roddier},
  \& {Close}}]{Graves1998}
{Graves}, J.~E., {Northcott}, M.~J., {Roddier}, F.~J., {Roddier}, C.~A., \&
  {Close}, L.~M. 1998, in Presented at the Society of Photo-Optical
  Instrumentation Engineers (SPIE) Conference, Vol. 3353, Proc. SPIE Vol. 3353,
  p. 34-43, Adaptive Optical System Technologies, Domenico Bonaccini; Robert K.
  Tyson; Eds., ed. D.~{Bonaccini} \& R.~K. {Tyson}, 34--43

\bibitem[{{Herbig}(1960)}]{Herbig1960}
{Herbig}, G.~H. 1960, \apjs, 4, 337

\bibitem[{{Hillenbrand}(1995)}]{Hillenbrand1995}
{Hillenbrand}, L.~A. 1995, PhD thesis, University of Massachusetts Amherst,
  Department of Physics and Astronomy

\bibitem[{{Hillenbrand} {et~al.}(1992){Hillenbrand}, {Strom}, {Vrba}, \&
  {Keene}}]{Hillenbrand1992}
{Hillenbrand}, L.~A., {Strom}, S.~E., {Vrba}, F.~J., \& {Keene}, J. 1992, \apj,
  397, 613

\bibitem[{{Lada} \& {Lada}(2003)}]{Lada2003}
{Lada}, C.~J. \& {Lada}, E.~A. 2003, \araa, 41, 57

\bibitem[{{Larson}(1979)}]{Larson1979}
{Larson}, R.~B. 1979, \mnras, 186, 479

\bibitem[{{Larson}(1981)}]{Larson1981}
---. 1981, \mnras, 194, 809

\bibitem[{{Levreault}(1988)}]{Levreault1988}
{Levreault}, R.~M. 1988, \apjs, 67, 283

\bibitem[{{Looney} {et~al.}(2000){Looney}, {Mundy}, \& {Welch}}]{Looney2000}
{Looney}, L.~W., {Mundy}, L.~G., \& {Welch}, W.~J. 2000, \apj, 529, 477

\bibitem[{{Looney} {et~al.}(2003){Looney}, {Mundy}, \& {Welch}}]{Looney2003}
---. 2003, \apj, 592, 255

\bibitem[{{Looney} {et~al.}(2006){Looney}, {Wang}, {Hamidouche}, {Safier}, \&
  {Klein}}]{Looney2006}
{Looney}, L.~W., {Wang}, S., {Hamidouche}, M., {Safier}, P.~N., \& {Klein}, R.
  2006, \apj, 642, 330

\bibitem[{{Lynds}(1962)}]{Lynds1962}
{Lynds}, B.~T. 1962, \apjs, 7, 1

\bibitem[{{Mac Low} \& {Klessen}(2004)}]{Mac2004}
{Mac Low}, M.-M. \& {Klessen}, R.~S. 2004, Reviews of Modern Physics, 76, 125

\bibitem[{{Miao} {et~al.}(1995){Miao}, {Mehringer}, {Kuan}, \&
  {Snyder}}]{Miao1995}
{Miao}, Y., {Mehringer}, D.~M., {Kuan}, Y.-J., \& {Snyder}, L.~E. 1995, \apjl,
  445, L59

\bibitem[{{Poetzel} {et~al.}(1992){Poetzel}, {Mundt}, \& {Ray}}]{Poetzel1992}
{Poetzel}, R., {Mundt}, R., \& {Ray}, T.~P. 1992, \aap, 262, 229

\bibitem[{{Ridge} {et~al.}(2003){Ridge}, {Wilson}, {Megeath}, {Allen}, \&
  {Myers}}]{Ridge2003}
{Ridge}, N.~A., {Wilson}, T.~L., {Megeath}, S.~T., {Allen}, L.~E., \& {Myers},
  P.~C. 2003, \aj, 126, 286

\bibitem[{{Rohlfs} \& {Wilson}(2000)}]{Rohlfs2000}
{Rohlfs}, K. \& {Wilson}, T.~L. 2000, {Tools of radio astronomy} (Tools of
  radio astronomy / K.~Rohlfs, T.L.~Wilson.~New York : Springer,
  2000.~(Astronomy and astrophysics library,ISSN0941-7834))

\bibitem[{{Saito} {et~al.}(2006){Saito}, {Saito}, {Moriguchi}, \&
  {Fukui}}]{Saito2006}
{Saito}, H., {Saito}, M., {Moriguchi}, Y., \& {Fukui}, Y. 2006, \pasj, 58, 343

\bibitem[{{Sault} {et~al.}(1995){Sault}, {Teuben}, \& {Wright}}]{Sault1995}
{Sault}, R.~J., {Teuben}, P.~J., \& {Wright}, M.~C.~H. 1995, in Astronomical
  Society of the Pacific Conference Series, Vol.~77, Astronomical Data Analysis
  Software and Systems IV, ed. R.~A. {Shaw}, H.~E. {Payne}, \& J.~J.~E.
  {Hayes}, 433--+

\bibitem[{{Shu} {et~al.}(1993){Shu}, {Najita}, {Galli}, {Ostriker}, \&
  {Lizano}}]{Shu1993}
{Shu}, F., {Najita}, J., {Galli}, D., {Ostriker}, E., \& {Lizano}, S. 1993, in
  Protostars and Planets III, ed. E.~H. {Levy} \& J.~I. {Lunine}, 3--45

\bibitem[{{Shu}(1977)}]{Shu1977}
{Shu}, F.~H. 1977, \apj, 214, 488

\bibitem[{{Tassis} \& {Mouschovias}(2005)}]{Tassis2005}
{Tassis}, K. \& {Mouschovias}, T.~C. 2005, \apj, 618, 769

\bibitem[{{Tobin} {et~al.}(2007){Tobin}, {Looney}, {Mundy}, {Kwon}, \&
  {Hamidouche}}]{Tobin2007}
{Tobin}, J.~J., {Looney}, L.~W., {Mundy}, L.~G., {Kwon}, W., \& {Hamidouche},
  M. 2007, \apj, 659, 1404

\bibitem[{{Wang} \& {Looney}(2007)}]{Wang2007}
{Wang}, S. \& {Looney}, L.~W. 2007, \apj, 659, 1360

\bibitem[{{Waters} \& {Waelkens}(1998)}]{Waters1998}
{Waters}, L.~B.~F.~M. \& {Waelkens}, C. 1998, \araa, 36, 233

\bibitem[{{Welch} {et~al.}(1996){Welch}, {Thornton}, {Plambeck}, {Wright},
  {Lugten}, {Urry}, {Fleming}, {Hoffman}, {Hudson}, {Lum}, {Forster}, {Thatte},
  {Zhang}, {Zivanovic}, {Snyder}, {Crutcher}, {Lo}, {Wakker}, {Stupar},
  {Sault}, {Miao}, {Rao}, {Wan}, {Dickel}, {Blitz}, {Vogel}, {Mundy},
  {Erickson}, {Teuben}, {Morgan}, {Helfer}, {Looney}, {de Gues}, {Grossman},
  {Howe}, {Pound}, \& {Regan}}]{Welch1996}
{Welch}, W.~J., {Thornton}, D.~D., {Plambeck}, R.~L., {Wright}, M.~C.~H.,
  {Lugten}, J., {Urry}, L., {Fleming}, M., {Hoffman}, W., {Hudson}, J., {Lum},
  W.~T., {Forster}, J.~.~R., {Thatte}, N., {Zhang}, X., {Zivanovic}, S.,
  {Snyder}, L., {Crutcher}, R., {Lo}, K.~Y., {Wakker}, B., {Stupar}, M.,
  {Sault}, R., {Miao}, Y., {Rao}, R., {Wan}, K., {Dickel}, H.~R., {Blitz}, L.,
  {Vogel}, S.~N., {Mundy}, L., {Erickson}, W., {Teuben}, P.~J., {Morgan}, J.,
  {Helfer}, T., {Looney}, L., {de Gues}, E., {Grossman}, A., {Howe}, J.~E.,
  {Pound}, M., \& {Regan}, M. 1996, \pasp, 108, 93

\bibitem[{{Williams} {et~al.}(1994){Williams}, {de Geus}, \&
  {Blitz}}]{Williams1994}
{Williams}, J.~P., {de Geus}, E.~J., \& {Blitz}, L. 1994, \apj, 428, 693

\bibitem[{{Wu} {et~al.}(2005){Wu}, {Zhang}, {Chen}, {Yang}, {Wei}, \&
  {Ho}}]{Wu2005}
{Wu}, Y., {Zhang}, Q., {Chen}, H., {Yang}, C., {Wei}, Y., \& {Ho}, P.~T.~P.
  2005, \aj, 129, 330

\bibitem[{{Yoshida} {et~al.}(1991){Yoshida}, {Kogure}, {Nakano}, {Tatematsu},
  \& {Wiramihardja}}]{Yoshida1991}
{Yoshida}, S., {Kogure}, T., {Nakano}, M., {Tatematsu}, K., \& {Wiramihardja},
  S.~D. 1991, \pasj, 43, 363

\bibitem[{{Yoshida} {et~al.}(1992){Yoshida}, {Kogure}, {Nakano}, {Tatematsu},
  \& {Wiramihardja}}]{Yoshida1992}
---. 1992, \pasj, 44, 77

\bibitem[{{Zinnecker} \& {Yorke}(2007)}]{Zinnecker2007}
{Zinnecker}, H. \& {Yorke}, H.~W. 2007, ArXiv e-prints, 707

\end{thebibliography}

\clearpage

\begin{deluxetable}{lccccrccrcc}
\tabletypesize{\scriptsize}
\tablewidth{0pt}
\tablecaption{Parameters of Identified CS Clumps}
\tablehead{
\colhead{Label} & \colhead{RA.}
& \colhead{Dec.} & \colhead{a$_o$} & \colhead{b$_o$} & \colhead{PA$_o$} 
& \colhead{a} & \colhead{b} & \colhead{PA}
& \colhead{R$_{eq}$} & \colhead{$\varepsilon$}\\
\colhead{} & \colhead{(J2000)} & \colhead{(J2000)}
& \colhead{$\arcsec$} & \colhead{$\arcsec$} & \colhead{degree}
& \colhead{$\arcsec$} & \colhead{$\arcsec$} & \colhead{degree}
&\colhead{10$^{-2}$ pc} &\colhead{} }
\startdata
A1 & 23:17:26.17 & 60:50:38.12 & 6.81$\pm$0.29 & 3.60$\pm$0.02 &-73.65$\pm$0.07 & 5.85 & 1.33 & -73.8&
1.49$\pm$0.04 & 0.97$\pm$0.04 \\
A2 & 23:17:26.78 & 60:50:38.75 & 11.77$\pm$1.37& 3.88$\pm$0.09 & 69.30$\pm$0.04 &11.26 & 1.85 &  69.1&
2.43$\pm$0.16 & 0.98$\pm$0.01 \\
A3 & 23:17:27.53 & 60:50:37.43 &             - &             - &              - & 3.49 & 3.34 & -67.6&
        1.82 &            - \\
A4 & 23:17:27.59 & 60:50:44.47 & 8.11$\pm$0.63 & 6.25$\pm$0.41 &-36.87$\pm$1.91 & 7.33 & 5.25 & -35.9&
3.31$\pm$0.19 & 0.69$\pm$0.09 \\
A5 & 23:17:27.55 & 60:50:50.40 &             - &             - &              - & 3.49 & 3.34 & -67.6&
        1.82 &            - \\
A6 & 23:17:28.11 & 60:50:51.31 & 6.75$\pm$0.77 & 4.95$\pm$0.13 &-11.99$\pm$0.82 & 5.84 & 3.55 & -10.7&
2.42$\pm$0.16 & 0.79$\pm$0.09 \\
A7 & 23:17:28.00 & 60:50:53.19 & 6.77$\pm$3.20 & 4.63$\pm$0.37 & -7.57$\pm$1.78 & 5.87 & 3.08 &  -6.5&
2.27$\pm$0.63 & 0.85$\pm$0.30 \\
A8 & 23:17:28.12 & 60:50:57.20 & 6.10$\pm$2.24 & 4.13$\pm$0.34 & 44.68$\pm$3.82 & 5.09 & 2.24 &  43.7&
1.80$\pm$0.42 & 0.89$\pm$0.23 \\
A9 & 23:17:28.55 & 60:50:57.32 & 5.73$\pm$0.89 & 3.31$\pm$0.07 &-81.44$\pm$0.32 & 3.49 & 3.34 & -67.6&
        1.82 &            - \\
B1 & 23:17:21.32 & 60:50:50.99 & 4.25$\pm$0.09 & 2.74$\pm$0.01 &-46.23$\pm$0.28 & 3.49 & 3.34 & -67.6&
        1.82 &            - \\
B2 & 23:17:21.84 & 60:50:49.41 & 6.81$\pm$3.35 & 4.49$\pm$0.49 & 87.86$\pm$1.67 & 5.86 & 2.97 &  87.0&
2.22$\pm$0.66 & 0.86$\pm$0.29 \\
B3 & 23:17:22.72 & 60:50:49.49 & 6.95$\pm$1.13 & 4.43$\pm$0.17 &-82.63$\pm$1.17 & 6.01 & 2.90 & -83.2&
2.23$\pm$0.22 & 0.87$\pm$0.08 \\
B4 & 23:17:23.25 & 60:50:46.58 & 5.16$\pm$0.87 & 3.35$\pm$0.29 & 78.28$\pm$1.41 & 3.49 & 3.34 & -67.6&
        1.82 &            - \\
B5 & 23:17:23.32 & 60:50:52.22 & 6.23$\pm$0.59 & 4.81$\pm$0.22 & 67.31$\pm$2.10 & 5.21 & 3.38 &  65.4&
2.24$\pm$0.14 & 0.76$\pm$0.09 \\
B6 & 23:17:24.79 & 60:50:45.95 & 6.98$\pm$3.60 & 3.72$\pm$0.11 &  0.47$\pm$0.40 & 6.12 & 1.33 &   1.1&
1.52$\pm$0.45 & 0.97$\pm$0.17 \\
B7 & 23:17:24.66 & 60:50:48.18 & 10.14$\pm$3.91& 4.98$\pm$0.08 &-19.60$\pm$0.19 & 9.55 & 3.62 & -19.2&
3.13$\pm$0.64 & 0.92$\pm$0.10 \\
B8 & 23:17:24.71 & 60:50:56.46 & 4.99$\pm$0.26 & 4.09$\pm$0.09 & 72.30$\pm$2.06 & 3.63 & 2.25 &  68.7&
1.52$\pm$0.06 & 0.78$\pm$0.06 \\
B9 & 23:17:25.20 & 60:50:58.36 &             - &             - &              - & 3.49 & 3.34 & -67.6&
        1.82 &            - \\
B10 & 23:17:24.58 & 60:51:02.37& 7.89$\pm$4.62 & 4.30$\pm$0.26 & 55.54$\pm$1.25 & 7.13 & 2.58 &  54.9&
2.28$\pm$0.75 & 0.93$\pm$0.20 \\
B11 & 23:17:24.98 & 60:51:02.42& 7.03$\pm$2.73 & 3.26$\pm$0.09 &-63.31$\pm$0.38 & 3.49 & 3.34 & -67.6&
        1.82 &            - \\
B12 & 23:17:24.45 & 60:51:04.92& 4.59$\pm$2.18 & 3.86$\pm$0.57 & 7.20$\pm$17.14 & 3.14 & 1.66 &   9.3&
1.22$\pm$0.47 & 0.84$\pm$0.65 \\
B13 & 23:17:24.81 & 60:51:08.29& 6.65$\pm$3.71 & 4.29$\pm$0.37 & 26.21$\pm$2.56 & 5.75 & 2.50 &  26.1&
2.02$\pm$0.66 & 0.90$\pm$0.30 \\
B14 & 23:17:25.15 & 60:51:10.18& 5.24$\pm$0.85 & 3.92$\pm$0.41 & 79.86$\pm$2.81 & 3.95 & 1.98 &  77.6&
1.49$\pm$0.22 & 0.86$\pm$0.16 \\
C1 & 23:17:27.90 & 60:50:27.43 &             - &             - &              - & 3.49 & 3.34 & -67.6&
        1.82 &            - \\
C2 & 23:17:28.39 & 60:50:27.61 &             - &             - &              - & 3.49 & 3.34 & -67.6&
        1.82 &            - \\
C3 & 23:17:28.06 & 60:50:23.99 &             - &             - &              - & 3.49 & 3.34 & -67.6&
        1.82 &            - \\
C4 & 23:17:27.69 & 60:50:22.08 &             - &             - &              - & 3.49 & 3.34 & -67.6&
        1.82 &            - \\
C5 & 23:17:28.85 & 60:50:09.20 & 5.49$\pm$0.19 & 4.01$\pm$0.05 &-58.20$\pm$0.70 & 4.24 & 2.22 & -57.5&
1.63$\pm$0.04 & 0.85$\pm$0.02 \\
D1 & 23:17:21.03 & 60:50:11.89 & 5.64$\pm$0.71 & 3.57$\pm$0.13 & 89.11$\pm$0.62 & 4.45 & 1.21 &  87.9&
1.24$\pm$0.12 & 0.96$\pm$0.06 \\
D2 & 23:17:20.40 & 60:50:12.78 & 6.36$\pm$0.97 & 4.76$\pm$0.29 &-59.04$\pm$3.04 & 5.31 & 3.38 & -58.5&
2.26$\pm$0.23 & 0.77$\pm$0.13 \\
D3 & 23:17:27.84 & 60:51:10.95 & 5.53$\pm$1.67 & 3.99$\pm$0.37 &-40.14$\pm$5.56 & 4.31 & 2.13 & -38.4&
1.61$\pm$0.34 & 0.86$\pm$0.23 \\
D4 & 23:17:27.60 & 60:51:12.97 & 4.57$\pm$1.09 & 3.88$\pm$0.55 &-43.04$\pm$27.76& 2.98 & 1.93 & -38.7&
1.28$\pm$0.29 & 0.76$\pm$0.38 \\
\enddata 
\end{deluxetable}

\clearpage

\begin{deluxetable}{lccc}
\tablewidth{0pt}
\tablecaption{Fitted Parameters of CS Integrated Spectra}
\tablehead{
\colhead{Label} 
& \colhead{Peak} & \colhead{Velocity}
& \colhead{FWHM} \\
\colhead{} &  \colhead{Jy}
& \colhead{km/s} & \colhead{km/s} }
\startdata
East & 4.79$\pm$0.31 & -31.68$\pm$0.06 & 1.87$\pm$0.14 \\
West(N) & 3.91$\pm$0.37 & -30.83$\pm$0.08 & 1.52$\pm$0.19 \\
West(S) & 1.73$\pm$0.25 & -34.13$\pm$0.26 & 3.24$\pm$0.69 \\
\enddata
\end{deluxetable}

\clearpage

\begin{deluxetable}{lccc}
\tablewidth{0pt}
\tablecaption{Spectral Fitting of CS Clumps}
\tablehead{
\colhead{Label}
& \colhead{Peak} & \colhead{Velocity}
& \colhead{FWHM}\\
\colhead{} & \colhead{Jy} & \colhead{km/s} & \colhead{km/s} }
\startdata
A1 & 0.44$\pm$0.05 & -31.42$\pm$0.07 & 1.30$\pm$0.16\\
A2 & 1.77$\pm$0.10 & -31.70$\pm$0.04 & 1.45$\pm$0.09\\
A3 & 0.68$\pm$0.08 & -32.35$\pm$0.05 & 0.85$\pm$0.11\\
A4 & 1.60$\pm$0.09 & -31.10$\pm$0.05 & 1.76$\pm$0.12\\
A5 & 0.45$\pm$0.05 & -31.37$\pm$0.10 & 1.63$\pm$0.23\\
A6 & 0.64$\pm$0.06 & -31.46$\pm$0.07 & 1.52$\pm$0.16\\
A7 & 0.54$\pm$0.05 & -31.31$\pm$0.08 & 1.70$\pm$0.20\\
A8 & 0.20$\pm$0.04 & -32.33$\pm$0.22 & 2.48$\pm$0.52\\
B2 & 0.36$\pm$0.07 & -30.65$\pm$0.09 & 1.00$\pm$0.22\\
B3 & 0.44$\pm$0.08 & -30.72$\pm$0.13 & 1.48$\pm$0.30\\
B6 & 0.50$\pm$0.04 & -30.92$\pm$0.06 & 1.55$\pm$0.14\\
B7 & 0.68$\pm$0.06 & -30.95$\pm$0.08 & 1.84$\pm$0.18\\
C1 & 0.28$\pm$0.04 & -30.81$\pm$0.13 & 1.76$\pm$0.31\\
C5 & 0.35$\pm$0.05 & -30.73$\pm$0.14 & 1.91$\pm$0.33\\
\enddata
\end{deluxetable}

\clearpage

\begin{deluxetable}{lcccccc}
\tablewidth{0pt}
\tablecaption{Parameters of the Observed Molecular Lines}
\tablehead{
\colhead{Molecular Name} 
& \colhead{Transition} & \colhead{$\nu$}
& \colhead{S$\mu^2$} & \colhead{Q} & \colhead{E$_u$} \\
\colhead{} & \colhead{} & \colhead{GHz}
& \colhead{Debye} & \colhead{} & \colhead{K} }
\startdata
CS & J $=$ 2 - 1 & 97.980968 & 7.71 
& 0.86 T$_{ex}$\tablenotemark{a} & 7.0 \\
$^{13}$CO & J $=$ 1 - 0 & 110.201370 & 0.01 
& 0.36 T$_{ex}$ & 5.3 \\
C$^{18}$O & J $=$ 1 - 0 & 109.782182 & 0.01 
& 0.36 T$_{ex}$ & 5.3 \\
\enddata
\tablenotetext{a}{T$_{ex}$ is the line excitation temperture.}
\tablerefs{\citet{Rohlfs2000}}
\end{deluxetable}

\clearpage

\begin{deluxetable}{lcccccc}
\tablewidth{0pt}
\tablecaption{Mass Estimation of CS Clumps}
\tablehead{
\colhead{Label} & \colhead{I$_t$} & \colhead{$\tau$}
& \colhead{M$_{LTE}$} & \colhead{N$_{lvg}$(CS)} 
& \colhead{M$_{LVG}$} & \colhead{M$_{vir}$} \\
\colhead{} & \colhead{Jy beam$^{-1}$ km s$^{-1}$}
& \colhead{} &  \colhead{M$_{\odot}$}
& \colhead{10$^{13}$ cm$^{-2}$} & \colhead{M$_{\odot}$}
& \colhead{M$_{\odot}$} }
\startdata
A1 & 1.755 & 0.46 & 3.59 & 3.43 - 5.56 & 0.64 -  1.04 & 2.38\\
A2 & 4.210 & 0.74 & 9.82 &      33.93 &        17.01 & 4.84\\
A3 & 1.631 & 0.34 & 3.15 & 3.64 - 5.91 & 1.02 -  1.65 & 1.24\\
A4 & 3.832 & 0.55 & 8.19 &       41.18 &        38.18 & 9.09\\
A5 & 1.557 & 0.38 & 3.08 & 4.30 - 6.98 & 1.20 -  1.95 & 4.57\\
A6 & 2.649 & 0.57 & 5.73 & 6.51 - 10.56& 3.25 -  5.26 & 5.31\\
A7 & 2.582 & 0.57 & 5.58 & 5.71 - 9.28 & 2.48 -  4.03 & 6.20\\
A8 & 1.328 & 0.33 & 2.56 & 2.48 - 6.55 & 0.68 -  1.80 & 10.47\\
A9 & 0.930 & 0.30 & 1.76 & - & - & -\\
B1 & 0.572 & 0.25 & 1.04 & - & - & -\\
B2 & 1.042 & 0.21 & 1.88 & 2.07 - 3.36 & 0.87 -  1.41 & 2.11\\\\
B3 & 1.399 & 0.31 & 2.67 & 3.91 - 6.33 & 1.64 -  2.66 & 4.62\\
B4 & 0.799 & 0.26 & 1.49 & - & - & -\\
B5 & 1.266 & 0.26 & 2.37 & - & - & -\\
B6 & 2.012 & 0.48 & 4.17 & 5.21 - 8.46 & 1.02 -  1.66 & 3.47\\
B7 & 4.457 & 0.72 &10.30 & 7.88 - 12.79& 6.56 - 10.65 & 10.05\\
B8 & 1.582 & 0.49 & 3.28 & - & - & -\\
B9 & 0.950 & 0.30 & 1.81 & - & - & -\\
B10 & 1.516& 0.30 & 2.88 & - & - & -\\
B11 & 1.031& 0.24 & 1.91 & - & - & -\\
B12 & 0.703& 0.22 & 1.28 & - & - & -\\
B13 & 1.102& 0.25 & 2.04 & - & - & -\\
B14 & 0.810& 0.24 & 1.55 & - & - & -\\
C1 & 1.103 & 0.33 & 2.12 & 2.85 - 5.91 & 0.80 -  1.65 & 5.33\\
C2 & 1.444 & 0.33 & 2.77 & - & - & -\\
C3 & 1.600 & 0.54 & 3.40 & - & - & -\\
C4 & 1.200 & 0.50 & 2.50 & - & - & -\\
C5 & 1.059 & 0.27 & 1.97 & 3.95 - 6.42 & 0.90 -  1.46 & 5.65\\
D1 & 0.914 & 0.23 & 1.66 & - & - & -\\
D2 & 1.188 & 0.24 & 2.20 & - & - & -\\
D3 & 0.878 & 0.24 & 1.61 & - & - & -\\
D4 & 0.782 & 0.26 & 1.45 & - & - & -\\
\enddata
\end{deluxetable}

\clearpage

\begin{figure}[ht]
\begin{center}
\includegraphics[width=0.5\textwidth]{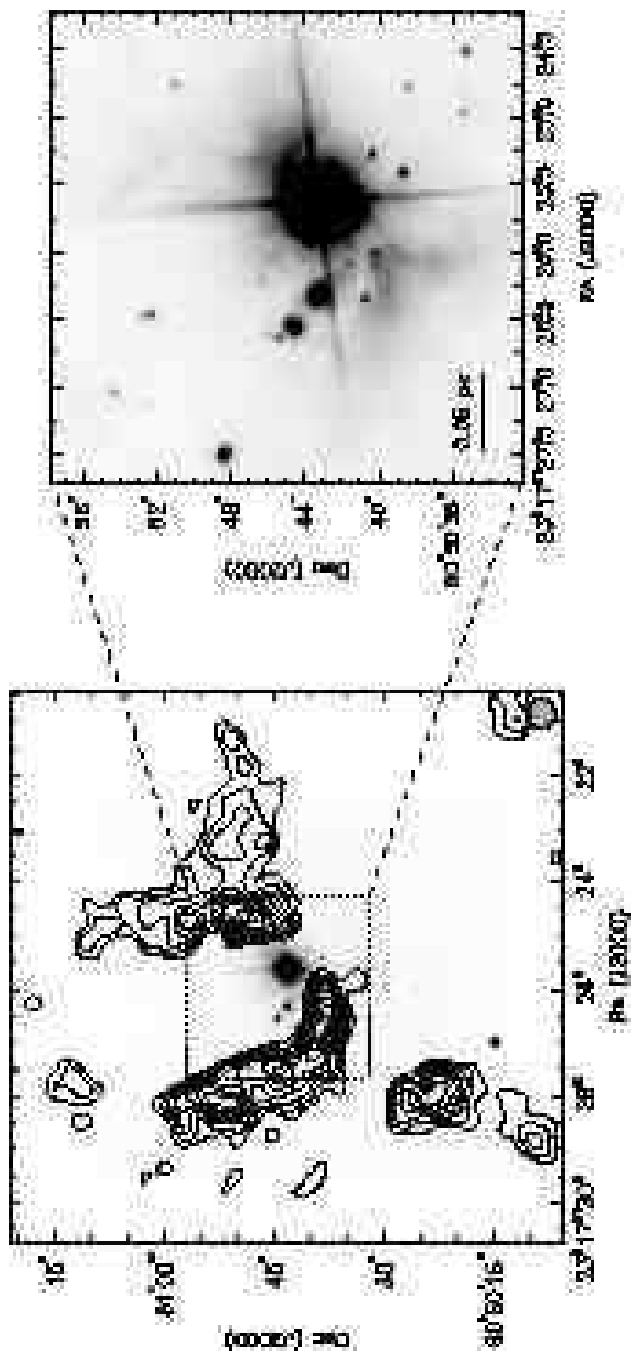}
\caption{Left: CS(2-1) emission toward the MWC 1080 system overlaid
on a small adaptive optics K$\arcmin$-band image taken from CFHT (Paper II).
The noise is 0.14 Jy/beam km/s. The contours are linearly
spaced from 2 to 10 times of noise. The beam, shown at the lower right hand corner, is 4$\farcs$04 $\times$
3$\farcs$84 with a PA of -62$\arcdeg$. 
Negative contours, mainly from resolved out large-scale emission,
are not shown here in order to simplify the image.
The dotted box indicated the zoomed field
of the K$\arcmin$-band image (right) in order to show the distribution
of young cluster members more clearly.}
\end{center}
\end{figure}

\begin{figure}[ht]
\begin{center}
\includegraphics[width=0.9\textwidth,angle=-90]{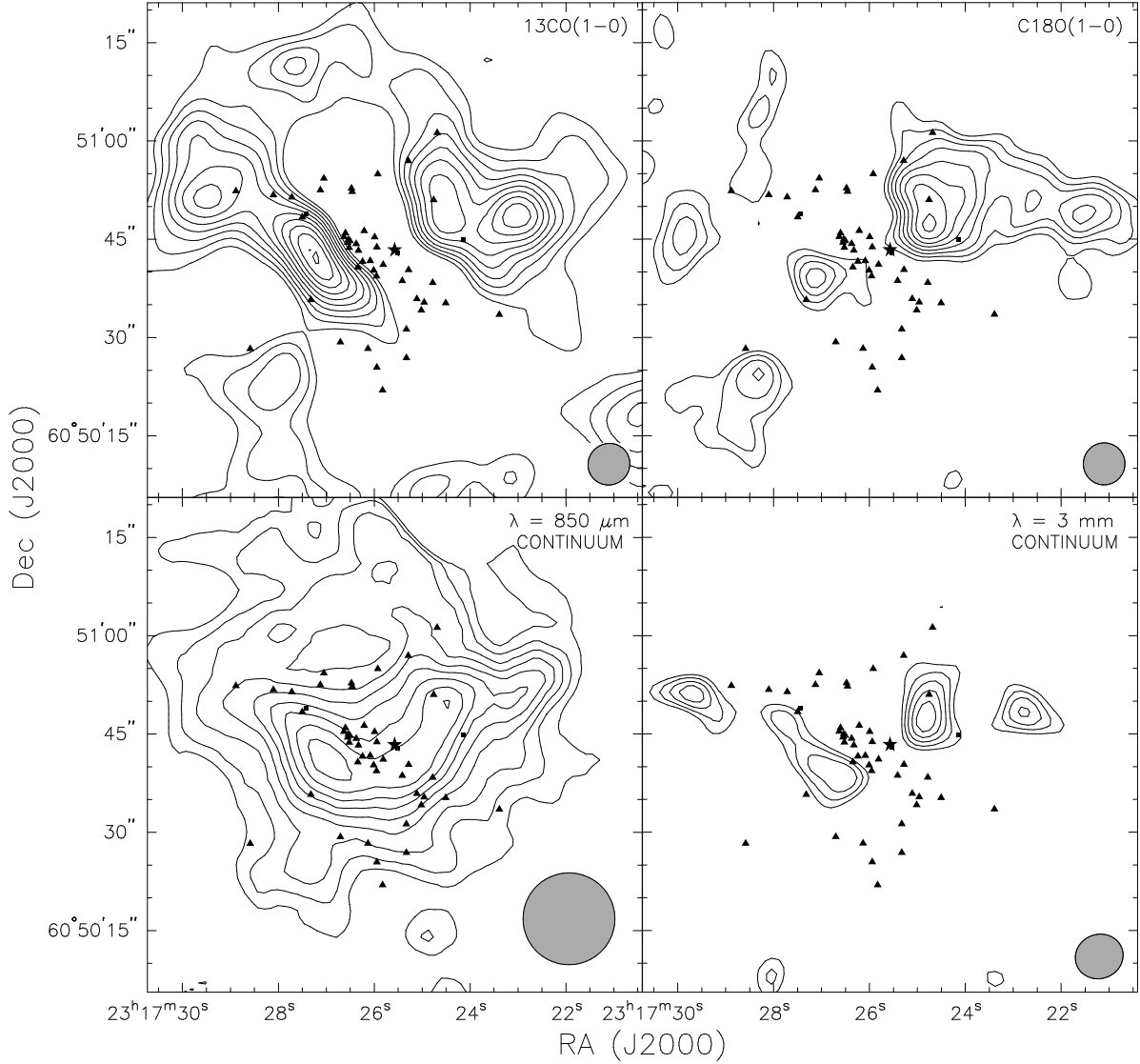}
\caption{$^{13}$CO, C$^{18}$O, and continuum emissions.
For $^{13}$CO, the noise is 0.3 Jy/beam km/s. The contours are linearly
spaced from 2 to 20 times of noise. The beam is 6$\farcs$44 $\times$
6$\farcs$28 with a PA of -51$\arcdeg$.
For C$^{18}$O, the noise is 0.12 Jy/beam km/s. The contours are linearly
spaced from 2 to 10 times of noise. The beam is 6$\farcs$41 $\times$
6$\farcs$34 with a PA of -13$\arcdeg$.
For 3 mm continuum map, the noise is 0.009 Jy/beam km/s. The contours are linearly
spaced from 2 to 6 times of noise. The beam is 7$\farcs$36 $\times$
6$\farcs$74 with a PA of -70$\arcdeg$. 
Beams are shown at the lower right hand corner of each panel.
The star symbols indicate the location of MWC 1080, and the triangle symbols indicate
the NIR-identified cluster members from Paper II.}
\end{center}
\end{figure}

\begin{figure}[ht]
\begin{center}
\includegraphics[width=0.9\textwidth,angle=-90]{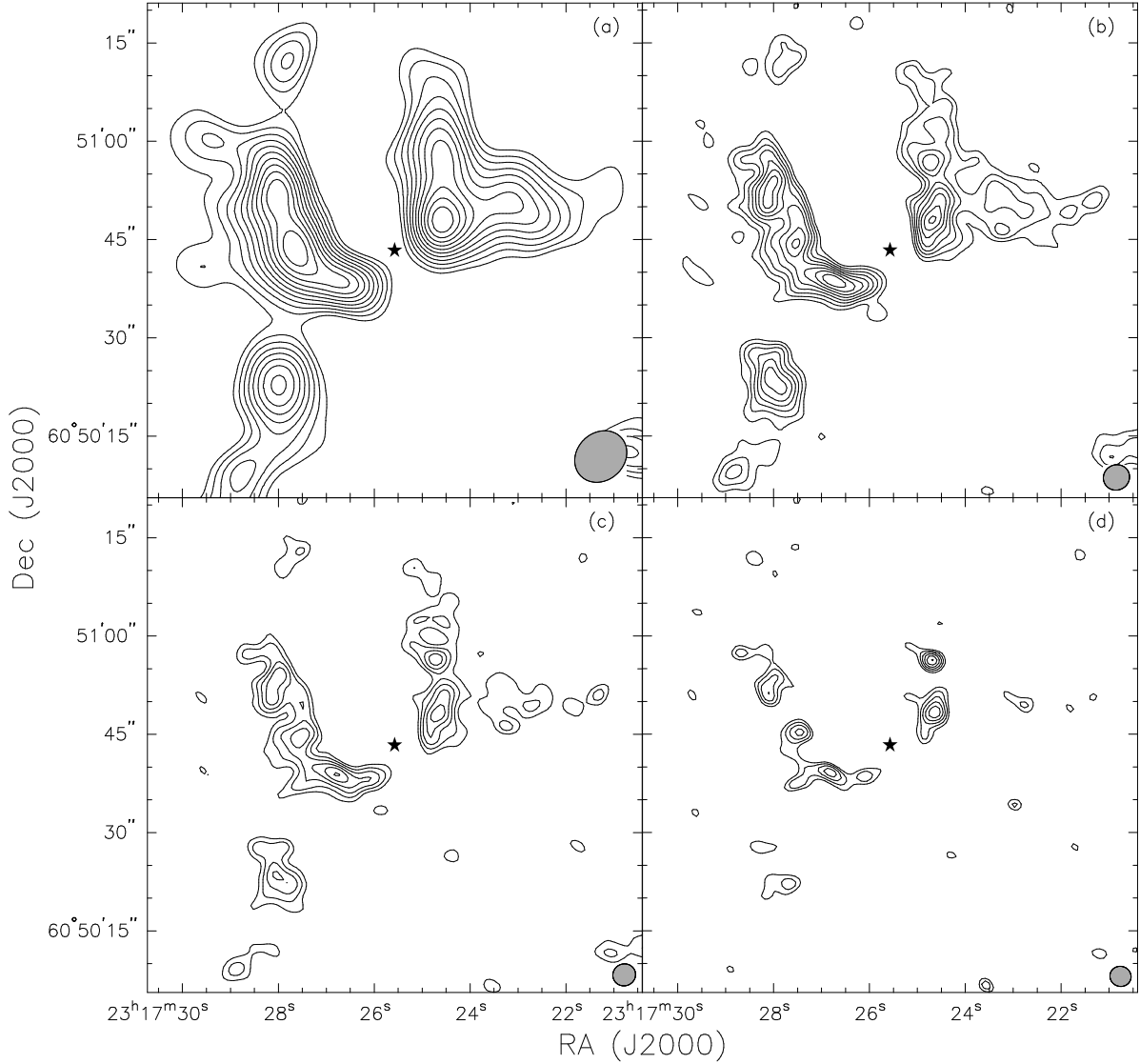}
\caption{The beam sizes are 8$\farcs$51 $\times$ 7$\farcs$25 with a PA of -46$\arcdeg$,
4$\farcs$04 $\times$ 3$\farcs$84 with a PA of -62$\arcdeg$, 3$\farcs$49 $\times$ 3$\farcs$34 with a PA of -67$\arcdeg$,
and 3$\farcs$24 $\times$ 3$\farcs$05 with a PA of 72$\arcdeg$, from (a) to (d) respectively.
Beams are shown at the lower right hand corner of each panel.
The contours are linearly spaced from (a) 2 to 13 times of noise,
0.23 Jy/beam km/s, (b) 2 to 10 times of noise, 0.14 Jy/beam km/s,
(c) 2 to 8 times of noise, 0.16 Jy/beam km/s,
(d) 2 to 5 times of noise, 0.25 Jy/beam km/s.
The star symbols indicate the location of MWC 1080.}

\end{center}
\end{figure}

\begin{figure}[ht]
\begin{center}
\includegraphics[width=0.8\textwidth,angle=-90]{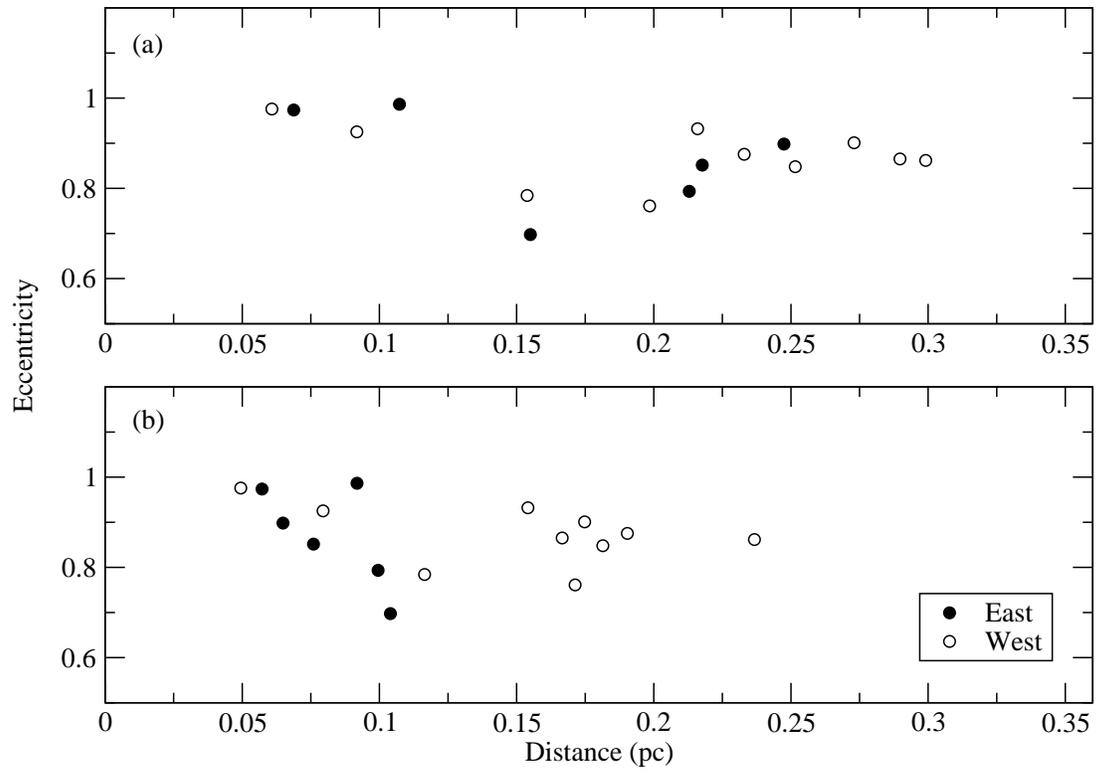}
\caption{The projected eccentricity of clumps vs. the distance to (a)
MWC 1080 and (b) outflow axis.} 
\end{center}
\end{figure}

\begin{figure}[ht]
\begin{center}
\includegraphics[width=0.8\textwidth,angle=-90]{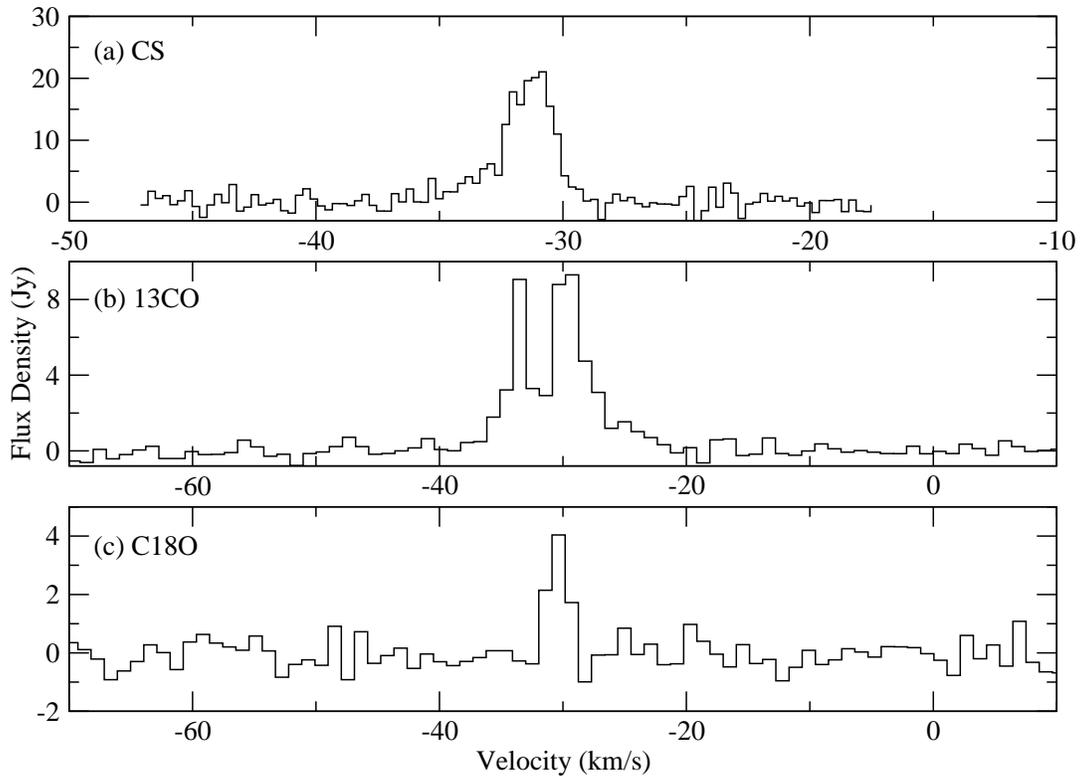}
\caption{Integrated spectra for overall gas traced by
CS, $^{13}$CO, and C$^{18}$O.}
\end{center}
\end{figure}

\begin{figure}[ht]
\begin{center}
\includegraphics[width=0.8\textwidth,angle=-90]{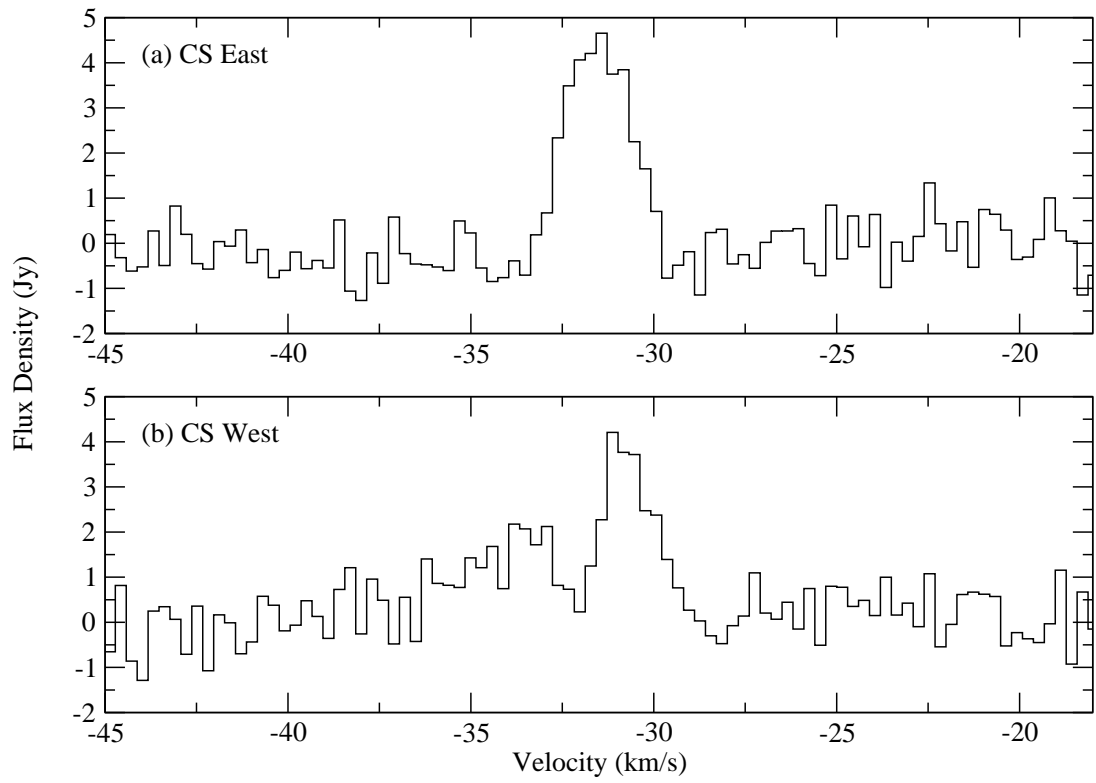}
\caption{Integrated spectra for CS East and West. 
East shows a broad linewidth; while West shows double-peaked
features. This double-peaked feature actually comes from 
two gas components with different velocities.}
\end{center}
\end{figure}

\begin{figure}[ht]
\begin{center}
\includegraphics[width=0.8\textwidth]{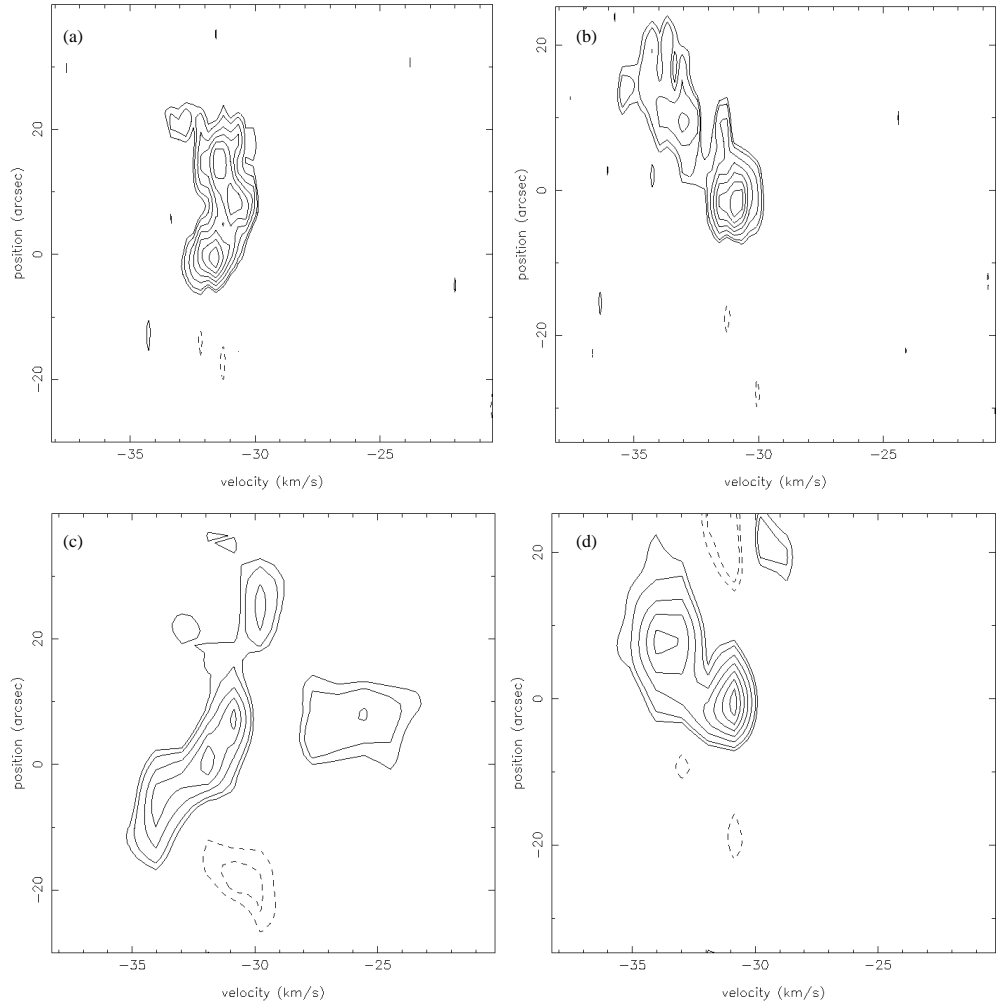}
\caption{Position-velocity (PV) diagrams along the outflow axis for (a) CS East,
(b) CS West, (c) $^{13}$CO East, and (d) $^{13}$CO West.
Contours are plotted from 1.5 $\sigma$, 2 $\sigma$, 3 $\sigma$ to maximum,
with $\sigma$ $=$ 0.1 Jy beam$^{-1}$.}
\end{center}
\end{figure}

\begin{figure}[ht]
\begin{center}
\includegraphics[width=0.8\textwidth]{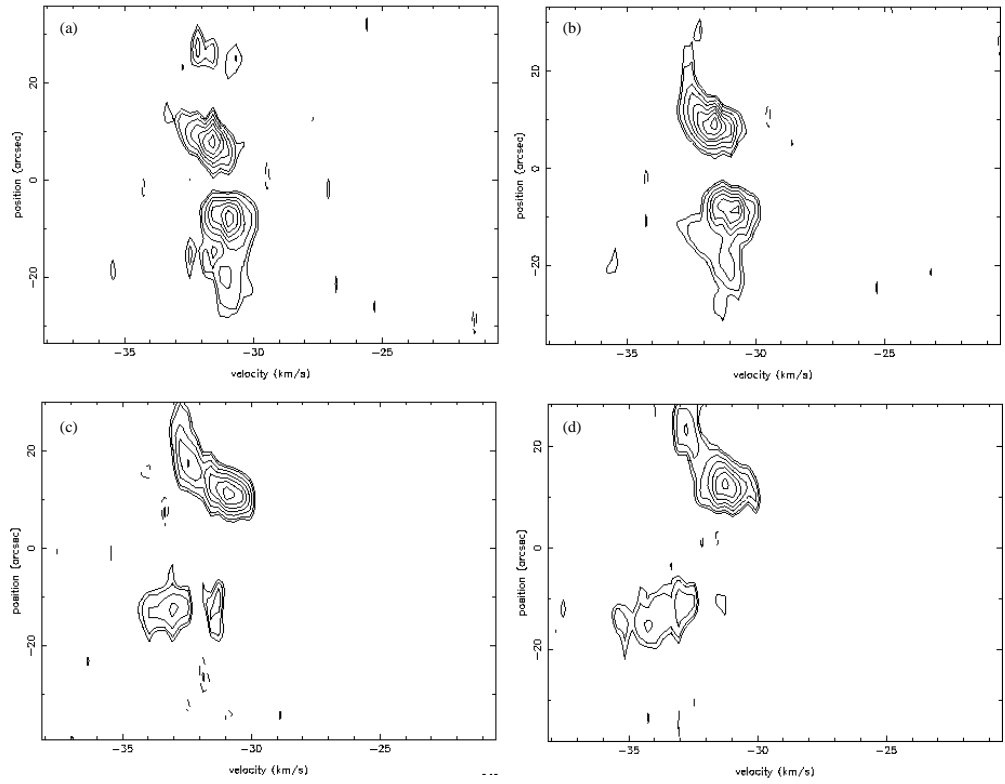}
\caption{PV diagrams along the hourglass axis from closer to
MWC 1080 (a) to farther to MWC 1080 (d).
Contours are plotted from 1.5 $\sigma$, 2 $\sigma$, 3 $\sigma$ to maximum,
with $\sigma$ $=$ 0.1 Jy beam$^{-1}$.}
\end{center}
\end{figure}

\begin{figure}[ht]
\begin{center}
\includegraphics[width=0.8\textwidth,angle=-90]{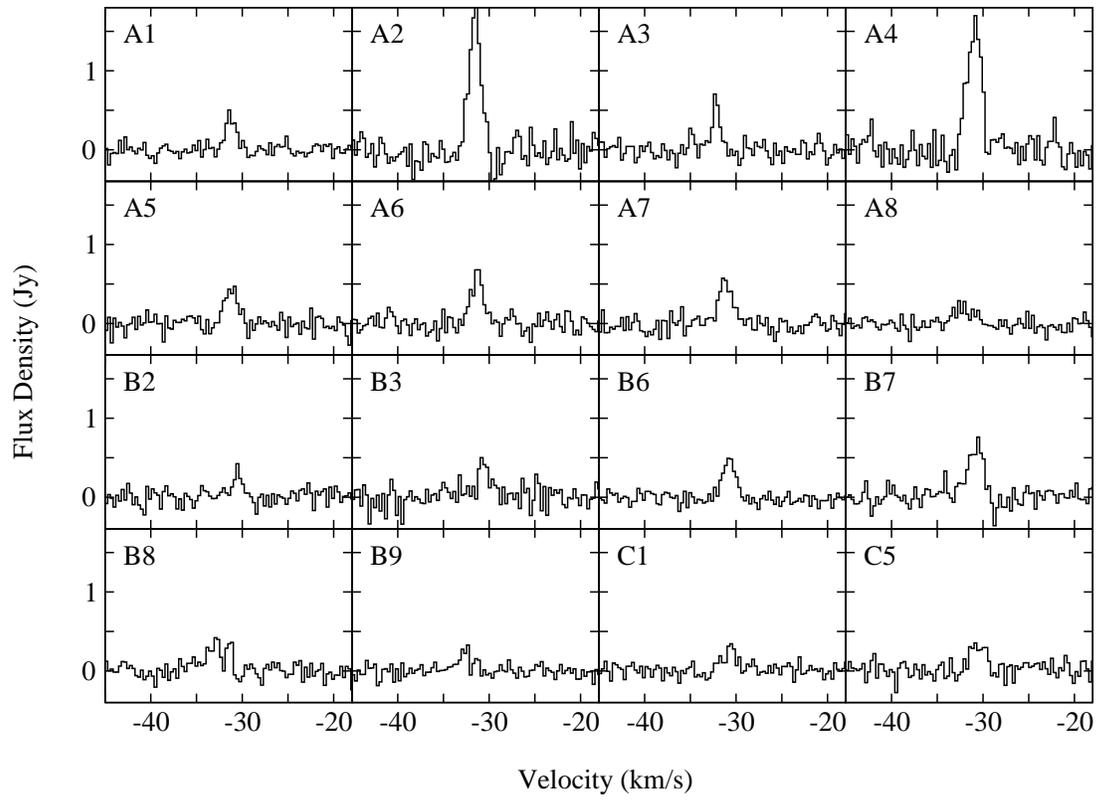}
\caption{Integrated CS spectra for identified CS clumps.}
\end{center}
\end{figure}

\begin{figure}[ht]
\begin{center}
\includegraphics[width=0.8\textwidth,angle=-90]{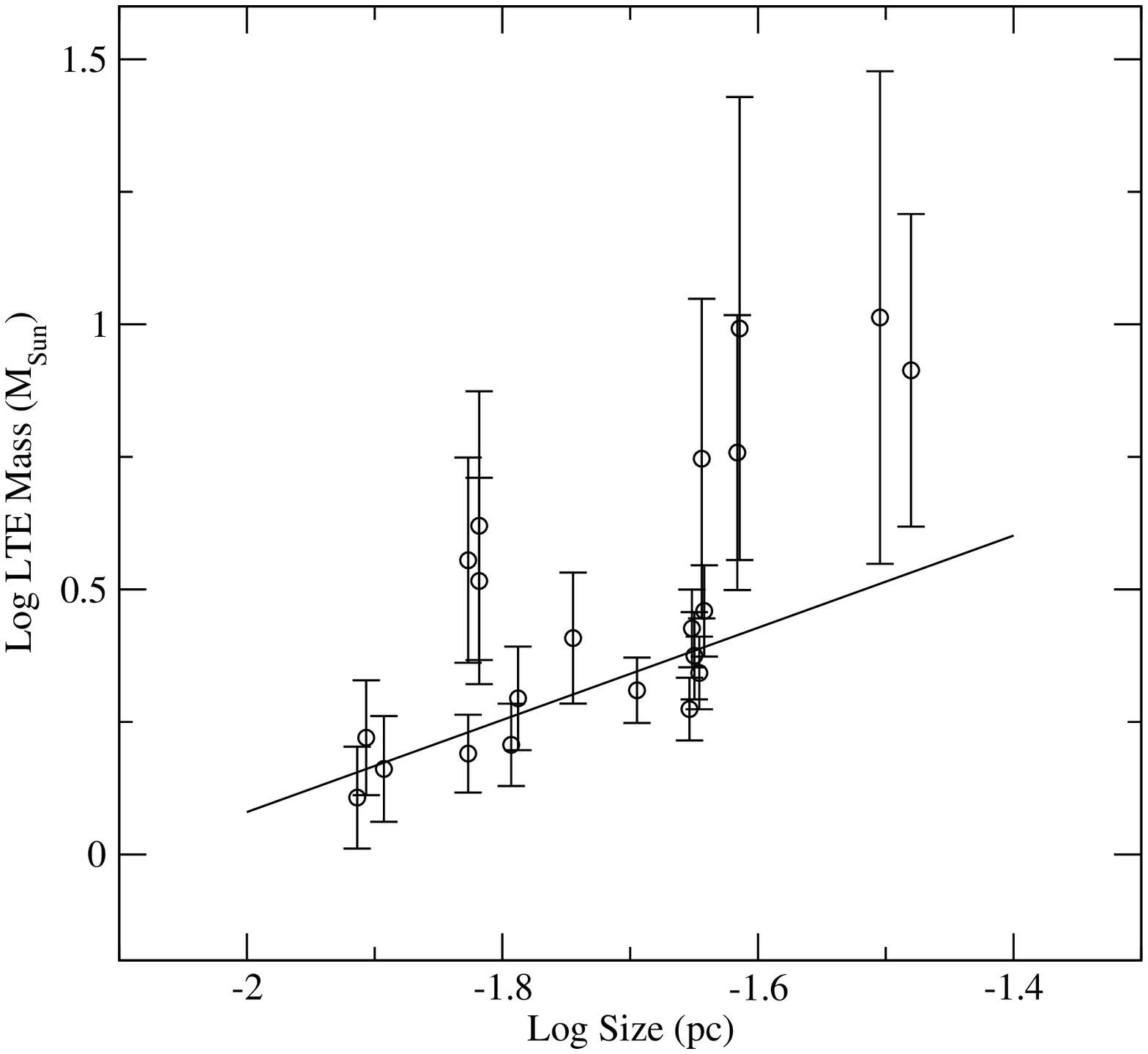}
\caption{Mass vs. size of clumps. The solid line is the fitted
relation.}
\end{center}
\end{figure}

\begin{figure}[ht]
\begin{center}
\includegraphics[width=0.8\textwidth,angle=-90]{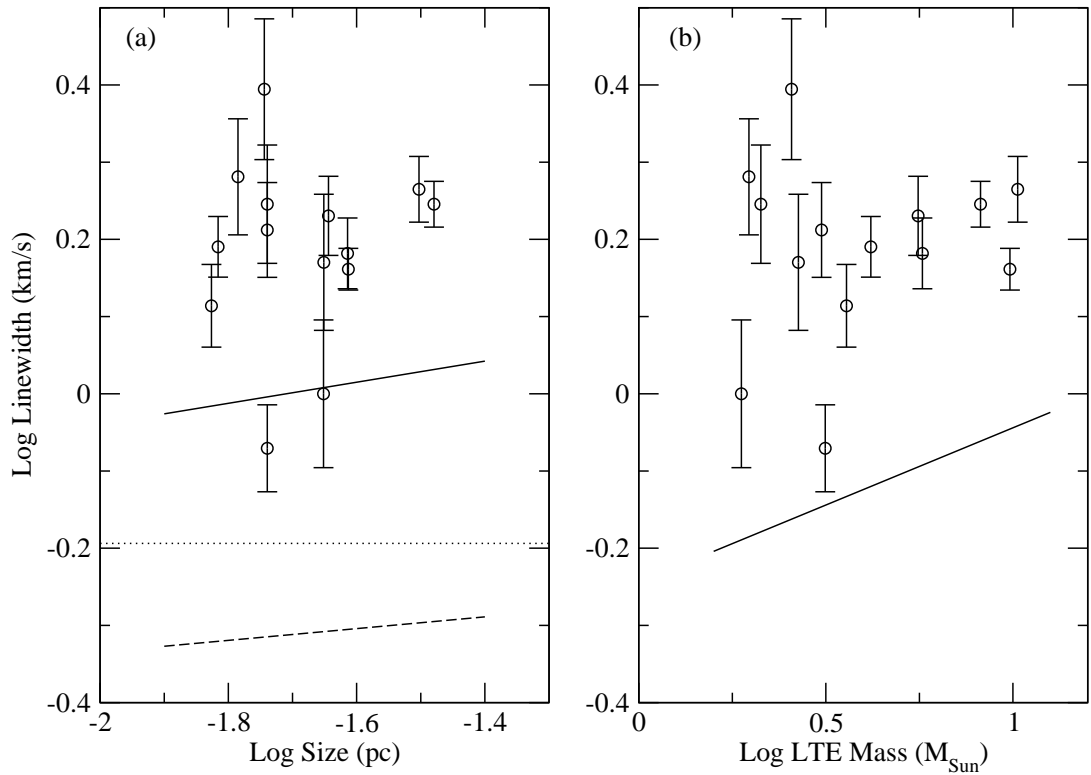}
\caption{The relationship between linewidth vs. 
(a) size and (b) LTE mass of CS clumps. 
The solid and dashed lines in (a) indicate the massive
cores and low-mass cores \citep[][]{Caselli1995}, respectively,
and the dotted line is the thermal motion at temperature of 20 K.
The solid line in (b) indicates the mass-linewidth relation
from \citet{Larson1981}.}
\end{center}
\end{figure}

\begin{figure}[ht]
\begin{center}
\includegraphics[width=0.8\textwidth,angle=-90]{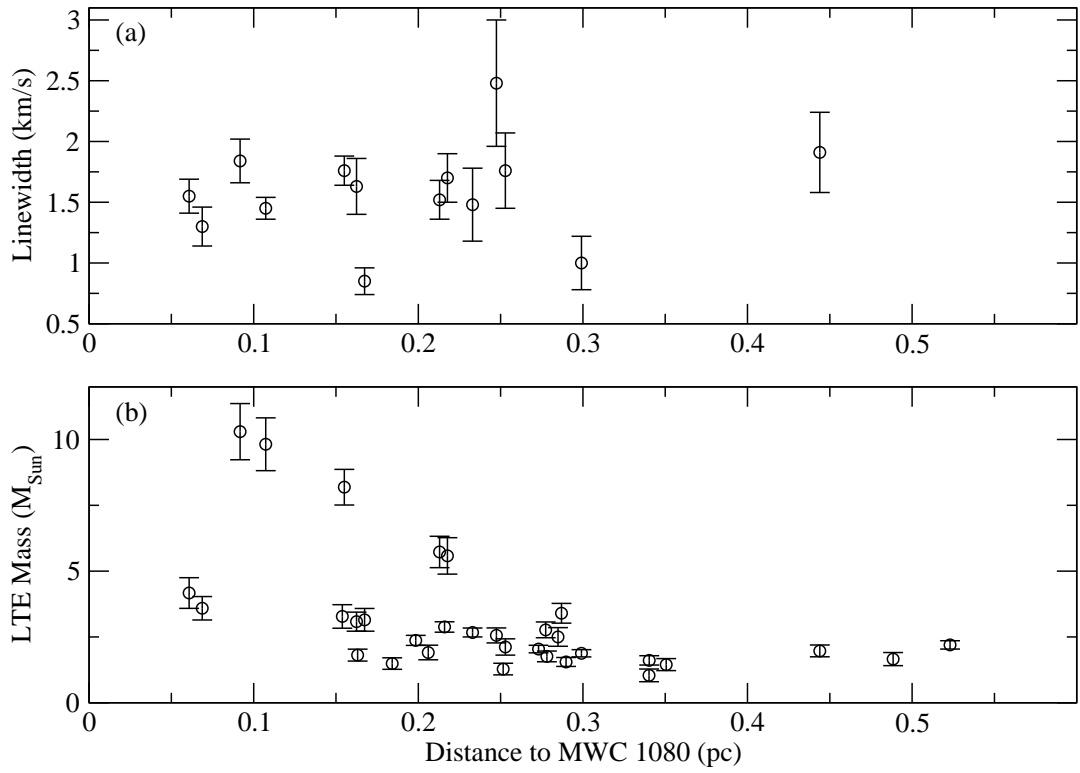}
\caption{(a) The CS linewidth and (b) the derived LTE mass vs. the distance
to MWC 1080.}
\end{center}
\end{figure}

\begin{figure}[ht]
\begin{center}
\includegraphics[width=0.8\textwidth,angle=-90]{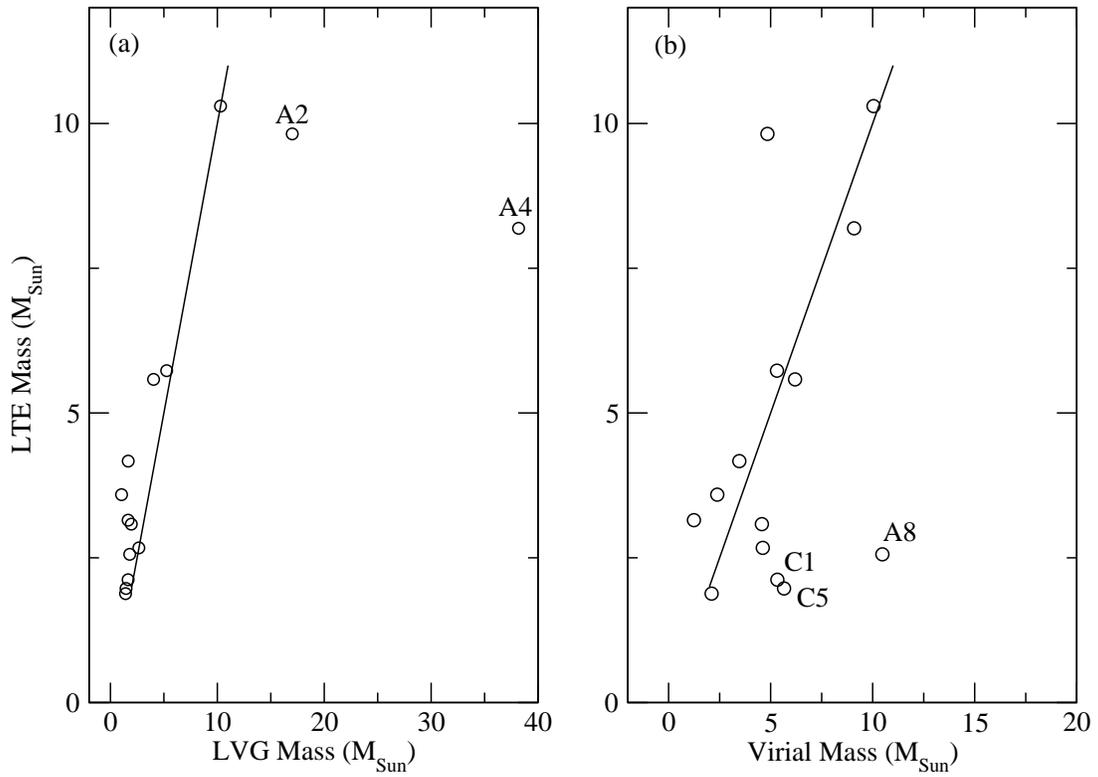}
\caption{The comparison between differently derived masses.
The solid lines are when the two masses are equal.}
\end{center}
\end{figure}

\end{document}